\documentclass[12pt]{article}
\usepackage{a4,amsmath,amsthm}
\usepackage{eucal}
\usepackage{amsfonts}
\newcommand{\ar}{\renewcommand{\arraystretch}{1}}
\DeclareMathAlphabet{\bb}{U}{msb}{m}{n}
\gdef\C{\Bbb C}
\gdef\dZ{\Bbb Z}

\gdef\K{\Bbb K}
\gdef\R{\Bbb R}
\gdef\BH{\Bbb H}
\gdef\F{\Bbb F}
\DeclareMathOperator{\End}{End}
\DeclareMathOperator{\spin}{{\bf Spin}}
\DeclareMathOperator{\pin}{{\bf Pin}}

\DeclareMathOperator{\Id}{Id}

\DeclareMathOperator{\Ker}{Ker}
\DeclareMathOperator{\bA}{{\bf A}}

\newcommand{\Om}{{\bf\Omega}} 
\newcommand{\cA}{\mathcal{A}}
\newcommand{\cB}{\mathcal{B}}

\newcommand{\cD}{\mathcal{D}}
\newcommand{\cE}{\mathcal{E}}

\newcommand{\M}{{\bf\sf M}}
\newcommand{\bsA}{{\bf\sf A}}
\newcommand{\sA}{{\sf A}}
\newcommand{\sB}{{\sf B}}
\newcommand{\sE}{{\sf E}}
\newcommand{\sC}{{\sf C}}
\newcommand{\sI}{{\sf I}}
\newcommand{\sW}{{\sf W}}
\newcommand{\bi}{{\bf i}}
\newcommand{\bj}{{\bf j}}
\newcommand{\bk}{{\bf k}}
\newcommand{\bx}{{\bf x}}

\newcommand{\bZ}{{\bf Z}}
\newcommand{\Lip}{\boldsymbol{\Gamma}}
\newcommand{\cl}{C\kern -0.2em \ell}

\newcommand{\p}{\prime}
\newcommand{\e}{\mbox{\bf e}}

\newtheorem{theorem}{Theorem}
\newtheorem{prop}{Proposition}
\begin{document}
\title{Fundamental Automorphisms of Clifford Algebras and an Extension
of D\c{a}browski Pin Groups}
\author{Vadim V. Varlamov\\
{\small\it Siberia State University of Industry, Novokuznetsk 654007, Russia}}
\date{}
\maketitle
\begin{abstract}
Double coverings of the orthogonal groups of the real and complex spaces
are considered. The relation between discrete transformations of these
spaces and fundamental automorphisms of Clifford algebras is established,
where an isomorphism between a finite group of the discrete transformations
and an automorphism group of the Clifford algebras plays a central role.
The complete classification of D\c{a}browski groups depending upon signatures
of the spaces is given. Two types of D\c{a}browski quotient groups are
introduced in case of odd-dimensional spaces. Application potentialities
of the introduced quotient groups in Physics are discussed.
\end{abstract}
\section{Introduction}
It is known that there are eight double coverings of the orthogonal group
$O(p,q)$ \cite{Dab88,BD89}:
\[
\rho^{a,b,c}:\;\pin^{a,b,c}(p,q)\longrightarrow O(p,q),
\]
where $a,b,c\in\{+,-\}$. The group $O(p,q)$ consists of four connected
components: identity connected component $O_0(p,q)$, and three components
corresponding to parity reversal $P$, time reversal $T$, and the
combination of these two $PT$, i.e., $O(p,q)=O_0(p,q)\cup P(O_0(p,q))\cup
T(O_0(p,q))\cup PT(O_0(p,q))$. Further, since the four element group
(reflection group) $\{1,P,T,PT\}$ is isomorphic to the finite group
$\dZ_2\otimes\dZ_2$ (Gauss-Klein group) \cite{Sal81a,Sal84}, then $O(p,q)$
may be represented by a semidirect product $O(p,q)\simeq O_0(p,q)\odot
(\dZ_2\otimes\dZ_2)$. 
The signs of $a,b,c$ correspond to the signs of the
squares of the elements in $\pin^{a,b,c}(p,q)$ which cover space reflection,
time reversal and a combination of these two: $P^2=a,\,T^2=b,\,(PT)^2=c$.
An explicit form of the group $\pin^{a,b,c}(p,q)$ is given by the following
semidirect product
\[
\pin^{a,b,c}(p,q)\simeq\frac{(\spin_0(p,q)\odot C^{a,b,c})}{\dZ_2},
\]
where $C^{a,b,c}$ are the four double coverings of $\dZ_2\otimes\dZ_2$.
The all eight double coverings of the orthogonal group $O(p,q)$ are given
in the following table:
\begin{center}{\renewcommand{\arraystretch}{1.3}
\begin{tabular}{|c|l|l|}\hline
$a$ $b$ $c$ & $C^{a,b,c}$ & Remark\\ \hline
$+$ $+$ $+$ & $\dZ_2\otimes\dZ_2\otimes\dZ_2$ & $PT=TP$\\
$+$ $-$ $-$ & $\dZ_2\otimes\dZ_4$ & $PT=TP$\\
$-$ $+$ $-$ & $\dZ_2\otimes\dZ_4$ & $PT=TP$\\
$-$ $-$ $+$ & $\dZ_2\otimes\dZ_4$ & $PT=TP$\\ \hline
$-$ $-$ $-$ & $Q_4$ & $PT=-TP$\\
$-$ $+$ $+$ & $D_4$ & $PT=-TP$\\
$+$ $-$ $+$ & $D_4$ & $PT=-TP$\\
$+$ $+$ $-$ & $D_4$ & $PT=-TP$\\ \hline
\end{tabular}
}
\end{center}
Here $Q_4$ is a quaternion group, and $D_4$ is a dihedral group. According
to \cite{Dab88} the group $\pin^{a,b,c}(p,q)$ satisfying to condition
$PT=-TP$ is called {\it Cliffordian}, and respectively {\it non-Cliffordian}
when $PT=TP$.

The $\pin$ and $\spin$ groups (Clifford-Lipschitz groups) widely used in
algebraic topology \cite{BH,Hae56,AtBSh,Kar68,Kar,KT89},
in the definition of pinor and spinor structures on the riemannian 
manifolds \cite{Mil63,Ger68,Ish78,Wh78,DT86,DP87,LM89,DR89,Cru91,AlCh94,
Ch94a,Ch94,CGT95,AlCh96},
spinor bundles \cite{RF90,RO90,FT96,Fr98,FT99},
and also have great importance in the theory of the Dirac operator on
manifolds \cite{Bau81,Bar91,Tr92,Fr97,Amm98}.
The Clifford-Lipschitz groups also intensively used in theoretical
physics \cite{CDD82,DW90,FRO90a,RS93,DWGK,Ch97}.
In essence, the D\c{a}browski group is a `detailed' (correct to a group of
discrete transformations) Clifford-Lipschitz group.

In the present paper the D\c{a}browski groups considered in the real and
complex spaces $\R^{p,q}$ and $\C^n$, respectively. The finite group
$\{1,P,T,PT\}$ is associated with an automorphism group $\{\Id,\star,
\widetilde{\phantom{cc}},\widetilde{\star}\}$ of the Clifford algebras
$\cl_{p,q}$ and $\C_{p+q}$. At first, consideration carried out for
even-dimensional algebras $\cl_{p,q},\,\C_{p+q}\,(p+q=2m)$ and associated
spaces $\R^{p,q}$ and $\C^{p+q}$. A relation between  signatures of the
Clifford algebras $(p,q)=(\underbrace{++\ldots+}_{p\,\text{times}},
\underbrace{--\ldots-}_{q\,\text{times}})$ and signatures of the
D\c{a}browski groups $(a,b,c)$ is established in section 4. It is shown
that there exist eight non-isomorphic relations between $(p,q)$ and
$(a,b,c)$ over the field $\F=\R$ and only two over the field $\F=\C$.
Further, odd-dimensional spaces are considered in section 5, at this point
the Clifford algebra is understood as a direct sum of two even-dimensional
subalgebras. It is shown (Theorem \ref{t11}) that in this case
there exist two quotient groups $\pin^b$ and $\pin^{b,c}$, the latter
group exists only over the field $\F=\C$: $\pin^{b,c}(p+q-1,\C)$ if
$p+q\equiv 1,5\pmod{8}$. The set of discrete transformations of the
quotient group $\pin^{b,c}(p+q-1,\C)$ does not form a group. It allows
to relate this group with some chiral field in Physics. In conclusion,
by way of example, the algebra $\C_3$ and the quotient group
$\pin^{b,c}(p+q-1,\C)$ are related with a Dirac-Hestenes spinor 
field \cite{Hest66,Hest90},
which has a broad application both in Physics and Geometry
\cite{Lou93,RVR93,RRSV95,RSVL,Var98a,Var98b}.
\section{Algebraic Preliminaries}
Clifford algebras play a key role in the definition of the $\pin$ groups.
Thus, in this section we will consider some basic facts about Clifford
algebras which relevant to definition and construction of the $\pin$ groups.
Let $\F$ be a field of characteristic 0 $(\F=\R,\,\F=\Om,\,\F=\C)$, where
$\Om$ is a field of double numbers $(\Om=\R\oplus\R)$, and $\R,\,\C$ are
the fields of real and complex numbers, respectively. A Clifford algebra
over a field $\F$ is an algebra with $2^n$ basis elements: $\e_0$
(unit of the algebra), $\e_1,\e_2,\ldots,\e_n$ and the products of the
one-index elements $\e_{i_1i_2\ldots i_k}=\e_{i_1}\e_{i_2}\ldots\e_{i_k}$.
Over the field $\F=\R$ the Clifford algebra is denoted as $\cl_{p,q}$, where
the indices $p,q$ correspond to the indices of the quadratic form
\[
Q=x^2_1+\ldots+x^2_p-\ldots-x^2_{p+q}
\]
of a vector space $V$ associated with $\cl_{p,q}$. A multiplication law
of $\cl_{p,q}$ is defined by the following rule:
\begin{equation}\label{e1}
\e^2_i=\sigma(q-i)\e_0,\quad\e_i\e_j=-\e_j\e_i,
\end{equation}
where 
\begin{equation}\label{e2}
\sigma(n)=\begin{cases}
-1 & \text{if $n\leq 0$},\\
+1 & \text{if $n>0$}.
\end{cases}
\end{equation}
The square of the volume element 
$\omega=\e_{12\ldots n}$, $n=p+q$,
plays an important role
in the theory of Clifford algebras:
\begin{equation}\label{e3}
\omega^2=\begin{cases}
-1 & \text{if $p-q\equiv 1,2,5,6\pmod{8}$},\\
+1 & \text{if $p-q\equiv 0,3,4,7\pmod{8}$}.
\end{cases}
\end{equation}  
A center $\bZ_{p,q}$ of $\cl_{p,q}$ consists of the unit $\e_0$ and the
volume element $\omega$. The element $\omega=\e_{12\ldots n}$ is belong
to a center when $n$ is odd. Indeed,
\begin{eqnarray}
\e_{12\ldots n}\e_i&=&(-1)^{n-i}\sigma(q-i)\e_{12\ldots i-1 i+1\ldots n},
\nonumber\\
\e_i\e_{12\ldots n}&=&(-1)^{i-1}\sigma(q-i)\e_{12\ldots i-1 i+1\ldots n},
\nonumber
\end{eqnarray}
therefore, $\omega\in\bZ_{p,q}$ if and only if $n-i\equiv i-1\pmod{2}$, that is,
$n$ is odd. Using (\ref{e3}) we have
\begin{equation}\label{e4}
\bZ_{p,q}=\begin{cases}
\phantom{1,}1 & \text{if $p-q\equiv 0,2,4,6\pmod{8}$},\\
1,\omega & \text{if $p-q\equiv 1,3,5,7\pmod{8}$}.
\end{cases}
\end{equation}
Moreover, when $n$ is odd a center $\bZ_{p,q}$ is isomorphic
to the fields $\C$ and $\Om$, respectively:
\[
\bZ_{p,q}\simeq\begin{cases}
\R\oplus i\R & \text{if $p-q\equiv 1,5\pmod{8}$},\\
\R\oplus e\R & \text{if $p-q\equiv 3,7\pmod{8}$},
\end{cases}
\]
where $e$ is a double unit $(e^2=1)$.

Further, let $\C_n=\C\otimes\cl_{p,q}$ and $\Om_{p,q}=\Om\otimes\cl_{p,q}$
be the Clifford algebras over the fields $\F=\C$ and $\F=\Om$, respectively.
\begin{theorem}\label{t1}
If $n=p+q$ is odd, then
\begin{eqnarray}
\cl_{p,q}&\simeq&\C_{p+q-1}\quad\text{if $p-q\equiv 1,5\pmod{8}$},\nonumber\\
\cl_{p,q}&\simeq&\Om_{p-1,q}\nonumber\\
&\simeq&\Om_{p,q-1}\quad\text{if $p-q\equiv 3,7\pmod{8}$}.\nonumber
\end{eqnarray}
\end{theorem}
\begin{proof} The structure of $\cl_{p,q}$ allows to identify the Clifford
algebras over the different fields. Indeed, transitions $\cl_{p-1,q}
\rightarrow\cl_{p,q},\;\cl_{p,q-1}\rightarrow\cl_{p,q}$ may be represented
as transitions from the real coordinates in $\cl_{p-1,q},\,\cl_{p,q-1}$
to complex coordinates of the form $a+\omega b$, where $\omega$ is
an additional basis element $\e_{12\ldots n}$ (volume element). Since
$n=p+q$ is odd, then the volume element $\omega$ in accordance with
(\ref{e4}) belongs to $\bZ_{p,q}$. Therefore, we can to identify it
with imaginary unit $i$ if $p-q\equiv 1,5\pmod{8}$ and with a double unit
$e$ if $p-q\equiv 3,7\pmod{8}$. The general element of the algebra $\cl_{p,q}$
has a form $\cA=\cA^{\p}+\omega\cA^{\p}$, where $\cA^{\p}$ is a general
element of the algebras $\cl_{p-1,q},\,\cl_{p,q-1}$.
\end{proof}
{\bf Example}. Let us consider the algebra $\cl_{0,3}$. According to the
theorem \ref{t1} we have $\cl_{0,3}\simeq\Om_{0,2}$, where $\Om_{0,2}$
is an algebra of elliptic biquaternions (it is a first so-called
Grassmann's extensive algebra introduced by Clifford in 1878 \cite{3}).
Since $\Om=\R\oplus\R$ and $\Om_{p,q}=\Om\otimes\cl_{p,q}$, we have
$\cl_{0,3}\simeq\cl_{0,2}\oplus\cl_{0,2}\simeq\BH\oplus\BH$, where $\BH$ is
a quaternion algebra.\\[0.4cm]
Generalizing this example we obtain
\begin{eqnarray}
\cl_{p,q}&\simeq&\cl_{p-1,q}\oplus\cl_{p-1,q}\nonumber\\
&\simeq&\cl_{p,q-1}\oplus\cl_{p,q-1}\quad\text{if $p-q\equiv 3,7\pmod{8}$}.
\label{e5'}
\end{eqnarray}

Over the field $\F=\C$ there is the analogous result \cite{Rash}.
\begin{theorem}\label{t2}
When $p+q\equiv 1,3,5,7\pmod{8}$ the Clifford algebra over the field
$\F=\C$ decomposes into a direct sum of two subalgebras:
\[
\C_{p+q}\simeq\C_{p+q-1}\oplus\C_{p+q-1}.
\]
\end{theorem}
In Clifford algebra $\cl_{p,q}$ 
there exist four fundamental
automorphisms \cite{Sch49,Rash}:

1) An automorphism $\cA\rightarrow\cA$.\\
This automorphism, obviously, is an identical automorphism of the algebra
$\cl_{p,q}$, $\cA$ is an arbitrary element of $\cl_{p,q}$.

2) An automorphism $\cA\rightarrow\cA^{\star}$.\\
In more details, for arbitrary element $\cA\in\cl_{p,q}$ there exists a
decomposition
\[
\cA=\cA^{\p}+\cA^{\p\p},
\]
where $\cA^{\p}$ is an element consisting of homogeneous odd elements, and
$\cA^{\p\p}$ is an element consisting of homogeneous even elements, 
respectively.
Then the automorphism $\cA\rightarrow\cA^{\star}$ is such that the element
$\cA^{\p\p}$ is not changed, and the element $\cA^{\p}$  changes sign:
\[
\cA^{\star}=-\cA^{\p}+\cA^{\p\p}.
\]
If $\cA$ is a homogeneous element, then
\begin{equation}\label{e5}
\cA^{\star}=(-1)^{k}\cA,
\end{equation}
where $k$ is a degree of element. It is easy to see that the automorphism
$\cA\rightarrow\cA^\star$ may be expressed via the volume element $\omega$:
\begin{equation}\label{e6}
\cA^\star=\omega\cA\omega^{-1},
\end{equation}
where $\omega^{-1}=(-1)^{\frac{n(n-1)}{2}}\omega$, $\cA$ is an arbitrary
element of $\cl_{p,q}$. When $k$ is odd, 
for the basis elements $\e_{i_1i_2\ldots
i_k}$ the sign changes, and when $k$ is even the sign is not changed. Over
the field $\F=\C$ we can multiply $\omega$ by $\varepsilon=\pm i^{\frac{n(n-1)}
{2}}$, then the equality (\ref{e6}) is not changed. At this point
we have always
$(\varepsilon\omega)^2=1$. Therefore,
\begin{equation}\label{e7}
\cA^\star=(\varepsilon\omega)\cA(\varepsilon\omega).
\end{equation}

3) An antiautomorphism $\cA\rightarrow\widetilde{\cA}$.\\
The antiautomorphism $\cA\rightarrow\widetilde{\cA}$ is a reversion
of the element $\cA$, that is, the substitution of the each basis element
$\e_{i_{1}i_{2}\ldots i_{k}}\in\cA$ by 
the element $\e_{i_{k}i_{k-1}\ldots i_{1}}$:
\[
\e_{i_{k}i_{k-1}\ldots i_{1}}=(-1)^{\frac{k(k-1)}{2}}\e_{i_{1}i_{2}\ldots
i_{k}}.
\]
Therefore, for any $\cA\in\cl_{p,q}$ we have
\begin{equation}\label{e8}
\widetilde{\cA}=(-1)^{\frac{k(k-1)}{2}}\cA.
\end{equation}

4) An antiautomorphism $\cA\rightarrow\widetilde{\cA^{\star}}$.\\
This antiautomorphism is a composition of the antiautomorphism
$\cA\rightarrow\widetilde{\cA}$ with the automorphism 
$\cA\rightarrow\cA^{\star}$. In the case of homogeneous element from
formulae (\ref{e5}) and (\ref{e8}) it follows
\begin{equation}\label{e9}
\widetilde{\cA^{\star}}=(-1)^{\frac{k(k+1)}{2}}\cA.
\end{equation}
It is obvious that\hspace{1mm} $\widetilde{\!\!\widetilde{\cA}}=\cA,\;
(\cA^{\star})^{\star}=
\cA,$ and $\widetilde{(\widetilde{\cA^{\star}})^{\star}}=\cA$.

The Lipschitz group $\Lip_{p,q}$, also called the Clifford group, introduced
by Lipschitz in 1886 \cite{Lips}, may be defined as the subgroup of
invertible elements $s$ of the algebra $\cl_{p,q}$:
\[
\Lip_{p,q}=\left\{s\in\cl^+_{p,q}\cup\cl^-_{p,q}\;|\;\forall x\in\R^{p,q},\;
s\bx s^{-1}\in\R^{p,q}\right\}.
\]
The set $\Lip^+_{p,q}=\Lip_{p,q}\cap\cl^+_{p,q}$ is called {\it special
Lipschitz group} \cite{Che1}.

Let $N:\;\cl_{p,q}\rightarrow\cl_{p,q},\;N(\bx)=\bx\widetilde{\bx}$.
If $\bx\in\R^{p,q}$, then $N(\bx)=\bx(-\bx)=-\bx^2=-Q(\bx)$. Further, the
group $\Lip_{p,q}$ has a subgroup
\begin{equation}\label{e10}
\pin(p,q)=\left\{s\in\Lip_{p,q}\;|\;N(s)=\pm 1\right\}.
\end{equation}
Analogously, {\it a spinor group} $\spin(p,q)$ is defined by the set
\begin{equation}\label{e11}
\spin(p,q)=\left\{s\in\Lip^+_{p,q}\;|\;N(s)=\pm 1\right\}.
\end{equation}
It is obvious that
\[
\spin(p,q)=\pin(p,q)\cap\cl^+_{p,q}.
\]
The group $\spin(p,q)$ contains a subgroup
\begin{equation}\label{e12}
\spin_+(p,q)=\left\{s\in\spin(p,q)\;|\;N(s)=1\right\}.
\end{equation}
It is easy to see that the groups $O(p,q),\,SO(p,q)$ and $SO_+(p,q)$ are
isomorphic correspondngly to the following quotient groups
\begin{eqnarray}
&&O(p,q)\simeq\pin(p,q)/\dZ_2,\nonumber\\
&&SO(p,q)\simeq\spin(p,q)/\dZ_2,\nonumber\\
&&SO_+(p,q)\simeq\spin_+(p,q)/\dZ_2,\nonumber
\end{eqnarray}
\begin{sloppypar}\noindent
where a kernel $\dZ_2=\{1,-1\}$. Thus, the groups $\pin(p,q)$, $\spin(p,q)$
and $\spin_+(p,q)$ are the double coverings of the groups $O(p,q),\,SO(p,q)$
and $SO_+(p,q)$, respectively.\end{sloppypar}

Further, since $\cl^+_{p,q}\simeq\cl^+_{q,p}$, then
\[
\spin(p,q)\simeq\spin(q,p).
\]
In contrast with this, the groups $\pin(p,q)$ and $\pin(q,p)$ are 
non-isomorphic. Denote $\spin(n)=\spin(n,0)\simeq\spin(0,n)$.
\begin{theorem}[{\rm\cite{Cor84}}]\label{t3}
The spinor groups
\[
\spin(2),\;\;\spin(3),\;\;\spin(4),\;\;\spin(5),\;\;\spin(6)
\]
are isomorphic to the unitary groups
\[
U(1),\;\;Sp(1)\sim SU(2),\;\;SU(2)\times SU(2),\;\;Sp(2),\;\;SU(4).
\]
\end{theorem} 
In accordance with Theorem \ref{t1} and decompositions (\ref{e5'})
over the field $\F=\R$ the algebra $\cl_{p,q}$ is isomorphic to a direct
sum of two mutually annihilating simple ideals $\frac{1}{2}(1\pm\omega)
\cl_{p,q}$: $\cl_{p,q}\simeq\frac{1}{2}(1+\omega)\cl_{p,q}\oplus\frac{1}{2}
(1-\omega)\cl_{p,q}$, where $\omega=\e_{12\ldots p+q},\,p-q\equiv 3,7
\pmod{8}$. At this point, each ideal is isomorpic to $\cl_{p-1,q}$ or
$\cl_{p,q-1}$. Therefore, for the Clifford-Lipschitz groups we have the
following isomorphisms
\begin{eqnarray}
\pin(p,q)&\simeq&\pin(p-1,q)\cup\pin(p-1,q)\nonumber\\
&\simeq&\pin(p,q-1)\cup\pin(p,q-1).
\end{eqnarray}
Or, since $\cl_{p-1,q}\simeq\cl^+_{p,q}\subset\cl_{p,q}$, then 
according to (\ref{e11})
\[
\pin(p,q)\simeq\spin(p,q)\cup\spin(p,q)
\]
if $p-q\equiv 3,7\pmod{8}$.

Further, when $p-q\equiv 1,5\pmod{8}$ from Theorem \ref{t1} it follows
that $\cl_{p,q}$ is isomorphic to a complex algebra $\C_{p+q-1}$. Therefore,
for the $\pin$ groups we obtain
\begin{eqnarray}
\pin(p,q)&\simeq&\pin(p-1,q)\cup\e_{12\ldots p+q}\pin(p-1,q)\nonumber\\
&\simeq&\pin(p,q-1)\cup\e_{12\ldots p+q}\pin(p,q-1)\label{e13}
\end{eqnarray}
if $p-q\equiv 1,5\pmod{8}$ and correspondingly
\begin{equation}\label{e14}
\pin(p,q)\simeq\spin(p,q)\cup\e_{12\ldots p+q}\spin(p,q).
\end{equation}
In case $p-q\equiv 3,7\pmod{8}$ we have isomorphisms which are analoguos
to (\ref{e13})-(\ref{e14}), since $\omega\cl_{p,q}\sim\cl_{p,q}$.
Generalizing we obtain the following
\begin{theorem}\label{t4}
Let $\pin(p,q)$ and $\spin(p,q)$ be the Clifford-Lipschitz groups of the
invertible elements of the algebras $\cl_{p,q}$ with odd dimensionality,
$p-q\equiv 1,3,5,7\pmod{8}$. Then
\begin{eqnarray}
\pin(p,q)&\simeq&\pin(p-1,q)\cup\omega\pin(p-1,q)\nonumber\\
&\simeq&\pin(p,q-1)\cup\omega\pin(p,q-1)\nonumber
\end{eqnarray}
and
\[
\pin(p,q)\simeq\spin(p,q)\cup\omega\spin(p,q),
\]
where $\omega=\e_{12\ldots p+q}$ is a volume element of $\cl_{p,q}$.
\end{theorem}
In case of low dimensionalities from Theorem \ref{t3} and Theorem
\ref{t4} it immediately follows
\begin{theorem}\label{t5}
For $p+q\leq 5$ and $p-q\equiv 3,5\pmod{8}$,
\begin{eqnarray}
\pin(3,0)&\simeq&SU(2)\cup iSU(2),\nonumber\\
\pin(0,3)&\simeq&SU(2)\cup eSU(2),\nonumber\\
\pin(5,0)&\simeq&Sp(2)\cup eSp(2),\nonumber\\
\pin(0,5)&\simeq&Sp(2)\cup iSp(2).\nonumber
\end{eqnarray}
\end{theorem}
\begin{proof}\begin{sloppypar}\noindent
Indeed, in accordance with Theorem \ref{t4} $\pin(3,0)\simeq\spin(3)
\cup\e_{123}\spin(3)$. Further, from Theorem \ref{t3} we have
$\spin(3)\simeq SU(2)$, and a square of the element $\omega=\e_{123}$ is
equal to $-1$, therefore $\omega\sim i$. Thus, $\pin(3,0)\simeq SU(2)\cup
iSU(2)$. For the group $\pin(0,3)$ a square of $\omega$ is equal to $+1$,
therefore $\pin(0,3)\simeq SU(2)\cup eSU(2)$, $e$ is a double unit.
As expected, $\pin(3,0)\not\simeq\pin(0,3)$. The isomorphisms for the
groups $\pin(5,0)$ and $\pin(0,5)$ are analogously proved.\end{sloppypar}
\end{proof}

Further, let $E$ be a vector space, then a homomorphism
\[
\rho:\;\cl_{p,q}\longrightarrow\End E,
\]
which maps the unit element of the algebra $\cl_{p,q}$ to $\Id_E$, is
called {\it a representation} of $\cl_{p,q}$ in $E$ ($\End E$ is an
endomorphism algebra of the space $E$). The dimensionality of $E$ is called
a degree of the representation. The addition in $E$ together with the
mapping $\cl_{p,q}\times E\rightarrow E,\;(a,x)\mapsto\rho(a)x,\,a\in\cl_{p,q},
\,x\in E$, turns $E$ in $\cl_{p,q}$-module, {\it a representation module}.
The representation $\rho$ is faithful if its kernel is zero, that is,
$\rho(a)x=0,\,\forall x \in E\Rightarrow a=0$. If the representation
$\rho$ has only two invariant subspaces $E$ and $\{0\}$, then $\rho$ is said
to be simple or irreducible. On the contrary case, $\rho$ is said to be
semi-simple, that is, it is a direct sum of simple modules, and in this case
$E$ is a direct sum of subspaces which are globally invariant under $\rho(a),
\;\forall a\in\cl_{p,q}$. The representation $\rho$ of $\cl_{p,q}$
induces a representation of the group $\pin(p,q)$ which we will denote
by the same symbol $\rho$, and also induces a representation of the group
$\spin(p,q)$ which we will denote by $\Delta_{p,q}$. In so doing, we have
the following \cite{Che1,Rash}
\begin{theorem}\label{t6}
If $p+q=2m$ and $p-q\equiv 0,2,4,6\pmod{8}$, then
\[
\cl_{p,q}\simeq\End_{\F}(I_{p,q})\simeq\M_{2^m}(\F)
\]
and
\[
\cl_{p,q}\simeq\End_{\F}(I_{p,q})\simeq\M_{2^m}(\F)\oplus\M_{2^m}(\F)
\]
if $p+q=2m+1$ and $p-q\equiv 1,3,5,7\pmod{8}$, where $\F=\R,\C,\Om,\BH$,
$I_{p,q}$ is a minimal left ideal of $\cl_{p,q}$, $\End_{\F}(I_{p,q})$ is
an algebra of linear transformations in $I_{p,q}$ over the field $\F$,
$\M_{2^m}(\F)$ is a matrix algebra.
\end{theorem}  

Let us consider matrix representations of the fundamental automorphisms
of $\cl_{p,q}$ over the field $\F$ when $p+q$ is even \cite{Sch49,Rash}.
We start with the antiautomorphism $\cA\rightarrow\widetilde{\cA}$. In
accordance with Theorem \ref{t6} in the matrix representation the
antiautomorphism $\cA\rightarrow\widetilde{\cA}$ corresponds to an
antiautomorphism of the matrix algebra $\M_{2^m}(\F)$:
\[
\sA\longrightarrow \sA^T,
\]
in virtue of the well-known relation $(\sA\sB)^T=\sB^T\sA^T$, where $T$ 
is a symbol
of transposition. On the other hand, in the matrix representation of the
elements $\cA\in\cl_{p,q}$ for the antiautomorphism $\cA\rightarrow
\widetilde{\cA}$ we have
\[
\sA\longrightarrow\widetilde{\sA}.
\]
The composition of the two antiautomorphisms $\sA^T\rightarrow \sA\rightarrow
\widetilde{\sA}$ gives an automorphism $\sA^T\rightarrow\widetilde{\sA}$
which is an internal automorphism of the algebra $M_{2^m}(\F)$:
\begin{equation}\label{e15}
\widetilde{\sA}=\sE\sA^T\sE^{-1},
\end{equation}
where $\sE$ is a matrix, by means of which the antiautomorphism
$\cA\rightarrow\widetilde{\cA}$ is expressed in the matrix representation
of the algebra $\cl_{p,q}$.

Further, for the automorphism $\cA\rightarrow\cA^\star$, defined
by the formula (\ref{e6}), in the matrix representation we have
\begin{equation}\label{e16}
\sA^\star=\sW\sA\sW^{-1},
\end{equation}
where $\sA$ is a matrix representing an arbitrary element of $\cl_{p,q}$,
$\sW$ is a matrix of the volume element $\omega=\e_{12\ldots n}$.
Over the field $\F=\C$ we can multiply $\sW$ by the factor
$\varepsilon=\pm i^{\frac{(p+q)(p+q-1)}{2}}$, then $(\varepsilon \sW)^2=1$.
Therefore, the relation (\ref{e16}) may be rewritten in the form
\begin{equation}\label{e17}
\sA^\star=\sW^{\p}\sA\sW^{\p},
\end{equation}
where $\sW^{\p}=\varepsilon \sW$.

Finally, for the antiautomorphism $\cA\rightarrow\widetilde{\cA^\star}$,
which is the composition of the antiautomorphism $\cA\rightarrow
\widetilde{\cA}$ with the automorphism $\cA\rightarrow\cA^\star$, using
(\ref{e15}) and (\ref{e17}) we obtain a following expression
\[
\widetilde{\sA^\star}=\sW^{\p}\sE\sA^T\sE^{-1}\sW^{\p},
\]
or 
\begin{equation}\label{e18}
\widetilde{\sA^\star}=(\sE{\sW^{\p}}^T)\sA^T(\sE{\sW^{\p}}^T)^{-1}.
\end{equation}
Denoting $\sE{\sW^{\p}}^T=\sC$, where $\sC$ is a matrix representation the
antiautomorphism $\cA\rightarrow\widetilde{\cA^\star}$, and substituting
into (\ref{e18}) we obtain definitely
\begin{equation}\label{e19}
\widetilde{\sA^\star}=\sC\sA^T\sC^{-1}.
\end{equation}
{\bf Example}. Let consider matrix representations of the fundamental
automorphisms of the Dirac algebra $\cl_{4,1}$. In virtue of Theorem
\ref{t1} there is an isomorphism $\cl_{4,1}\simeq\C_4$, and therefore
$\cl_{4,1}\simeq\C_4\simeq\M_4(\C)$. In the capacity of the matrix
representations of the units $\e_i\in\C_4$ $(i=1,2,3,4)$ we take the
well-known Dirac $\gamma$-matrices (so-called canonical representation):
\[
\ar
\gamma_1=\begin{pmatrix}
0 & 0 & 0 & -i\\
0 & 0 & -i& 0\\
0 & i & 0 & 0\\
i & 0 & 0 & 0
\end{pmatrix},\quad\gamma_2=\begin{pmatrix}
0 & 0 & 0 & -1\\
0 & 0 & 1 & 0\\
0 & 1 & 0 & 0\\
-1& 0 & 0 & 0
\end{pmatrix},
\]
\begin{equation}\label{e20}\ar
\gamma_3=\begin{pmatrix}
0 & 0 & -i & 0\\
0 & 0 & 0 & i\\
i & 0 & 0 & 0\\
0 & -i& 0 & 0
\end{pmatrix},\quad\gamma_4=\begin{pmatrix}
1 & 0 & 0 & 0\\
0 & 1 & 0 & 0\\
0 & 0 &-1 & 0\\
0 & 0 & 0 &-1
\end{pmatrix}.
\end{equation}   
$\gamma$-matrices form the only one basis from the set of isomorphic
matrix basises of $\cl_{4,1}\simeq\C_4$. In the basis (\ref{e20}) the element
$\omega=\e_1\e_2\e_3\e_4$ is represented by a matrix $\sW=\gamma_5=\gamma_1
\gamma_2\gamma_3\gamma_4$. Since $\gamma^2_5=1$, then $\varepsilon=1\;
(\sW^{\p}=\sW)$ and the matrix
\begin{equation}\label{e21}\ar
\sW^T=\sW=\begin{pmatrix}
0 & 0 &-1 & 0\\
0 & 0 & 0 &-1\\
-1& 0 & 0 & 0\\
0 &-1 & 0 & 0
\end{pmatrix}
\end{equation}
in accordance with (\ref{e17}) is a matrix of the automorphism $\cA
\rightarrow\cA^\star$. Further, in the matrix representation the
antiautomorphism $\cA\rightarrow\widetilde{\cA}$ is defined by the
transformation $\widetilde{\sA}=\sE\sA^T\sE^{-1}$. For the $\gamma$-matrices
 we have
$\gamma^T_1=-\gamma_1,\,\gamma^T_2=\gamma_2,\,\gamma^T_3=-\gamma_3,
\,\gamma^T_4=\gamma_4$. Further,
\begin{eqnarray}
&&\gamma_1=-\sE\gamma_1\sE^{-1},\quad\gamma_2=\sE\gamma_2\sE^{-1},\nonumber\\
&&\gamma_3=-\sE\gamma_3\sE^{-1},\quad\gamma_4=\sE\gamma_4\sE^{-1}.\nonumber
\end{eqnarray}
It is easy to verify that a matrix $\sE=\gamma_1\gamma_3$ satisfies the
latter relations and, therefore, the antiautomorphism $\cA\rightarrow
\widetilde{\cA}$ in the basis (\ref{e20}) is defined by the matrix
\begin{equation}\label{e22}\ar
\sE=\gamma_1\gamma_3=\begin{pmatrix}
0 & -1 & 0 & 0\\
1 & 0 & 0 & 0\\
0 & 0 & 0 & -1\\
0 & 0 & 1 & 0
\end{pmatrix}.
\end{equation}
Finally, for the matrix $\sC=\sE\sW^T$ of the antiautomorphism $\cA\rightarrow
\widetilde{\cA^\star}$ from (\ref{e21}) and (\ref{e22}) in accordance with
(\ref{e19}) we obtain
\begin{equation}\label{e23}\ar
\sC=\sE\sW^T=\begin{pmatrix}
0 & 0 & 0 & 1\\
0 & 0 &-1 & 0\\
0 & 1 & 0 & 0\\
-1& 0 & 0 & 0
\end{pmatrix}.
\end{equation}
\section{Fundamental Automorphisms of\protect\newline 
Odd-dimensional Clifford Algebras}
Let us consider the fundamental automorphisms of the algebras $\cl_{p,q}$
and $\C_{p+q}$, where $p-q\equiv 1,3,5,7\pmod{8}$, $p+q=2m+1$. In accordance
with (\ref{e5'}) and Theorem \ref{t2} the algebras $\cl_{p,q}$ and
$\C_{p+q}$ are isomorphic to direct sums of two algebras with the even
dimensionality if correspondingly $p-q\equiv 3,7\pmod{8}$ and $p+q\equiv 1,3,5,7
\pmod{8}$. Therefore, matrix representations of $\cl_{p,q},\,\C_{p+q}$
are isomorphic to the direct sums of complete matrix algebras
$\M_{2^m}(\F)\oplus\M_{2^m}(\F)$, here $\F=\R,\,\F=\C$. On the other hand,
there exists an homomorphic mapping of $\cl_{p,q}$ and $\C_{p+q}$ into
one full matrix algebra $\M_{2^m}(\F)$ with preservation of addition,
multiplication and multiplication by the number. Besides, in the case
of $\F=\R$ and $p-q\equiv 1,5\pmod{8}$ the algebra $\cl_{p,q}$ is isomorphic
to the full matrix algebra $\M_{2^m}(\C)$ (Theorem \ref{t1}), therefore,
representations of the fundamental automorphisms of this algebra may be
realized by means of $\M_{2^m}(\C)$.
\begin{theorem}\label{t7}
If $p+q=2m+1$, then the following homomorphisms take place\\
1) $\F=\R$
\[
\epsilon:\;\;\cl_{p,q}\longrightarrow\M_{2^m}(\R)\quad\text{if}\;\;
p-q\equiv 3,7\pmod{8}.
\]
2) $\F=\C$
\[
\epsilon^{\p}:\;\;\C_{p+q}\longrightarrow\M_{2^m}(\C)\quad\text{if}\;\;
p+q\equiv 1,3,5,7\pmod{8}.
\]
\end{theorem}
\begin{proof} We start the proof with a more general case of $\F=\C$. 
According to (\ref{e4}) the volume element $\omega$ belongs to a center
of $\C_{n}$ ($n=p+q$), therefore, $\omega$ commutes with all basis elements
of this algebra and $(\varepsilon\omega)^2=1$. Further, recalling that a
vector complex space $\C^n$ is associated with the algebra $\C_n$, we see
that basis vectors $\{e_{1},e_{2},\ldots,e_{n}\}$ generate a 
subspace $C_{n}\subset C_{n+1}$. Thus, the algebra $\C_{n}$ in $C_{n}$ is
a subalgebra of $\C_{n+1}$ and consists of the elements which does not contain
the element $\e_{n+1}$.
A decomposition of the each element $\cA\in\C_{n+1}$ may be written in the
form
\[
\cA=\cA^{1}+\cA^{0},
\]
where $\cA^{0}$ is a set of all elements which contain $\e_{n+1}$, and
$\cA^{1}$ is a set of all elements which does not contain $\e_{n+1}$, therefore
$\cA^{1}\in\C_{n}$. If multiply $\cA^{0}$ by $\varepsilon\omega$, then the
elements $\e_{n+1}$ are mutually annihilate, therefore $\varepsilon\omega
\cA^{0}\in\C_{n}$. Denoting $\cA^{2}=\varepsilon\omega\cA^{0}$
and taking into account $(\varepsilon\omega)^2=1$  we obtain
\[
\cA=\cA^{1}+\varepsilon\omega\cA^{2},
\]
where $\cA^{1},\,\cA^{2}\in\C_{n}$. Consider now an homomorphism
$\epsilon:\;\C_{n+1}\rightarrow\C_{n}$, an action of which is defined by
the following law
\begin{equation}\label{e24}
\epsilon:\;\cA^{1}+\varepsilon\omega\cA^{2}\longrightarrow\cA^{1}+\cA^{2}.
\end{equation}
Obviously, at this point the all operations (addition, multiplication, and
multiplication by the number) are preserved. Indeed, let
\[
\cA=\cA^{1}+\varepsilon\omega\cA^{2},\quad
\cB=\cB^{1}+\varepsilon\omega\cB^{2},
\]
then in virtue of $(\varepsilon\omega)^{2}=1$ and 
commutativity of $\omega$ with all elements, we have for multiplication
\begin{multline}
\cA\cB=(\cA^{1}\cB^{1}+\cA^{2}\cB^{2})+\varepsilon\omega(\cA^{1}\cB^{2}+
\cA^{2}\cB^{1})\stackrel{\epsilon}{\longrightarrow}\\
(\cA^{1}\cB^{1}+
\cA^{2}\cB^{2})+(\cA^{1}\cB^{2}+\cA^{2}\cB^{1})=
(\cA^{1}+\cA^{2})(\cB^{1}+\cB^{2}).\nonumber
\end{multline}
that is, the image of product equals to the product of factor images in
the same order.

In the particular case of $\cA=\varepsilon\omega$ we have $\cA^{1}=0$ and
$\cA^{2}=1$, therefore
\[
\varepsilon\omega\longrightarrow 1.
\]
Thus, a kernel of the homomorphism $\epsilon$ consists of all elements
of the form $\cA^{1}-\varepsilon\omega\cA^{1}$, which under action of 
$\epsilon$ are mapped into
zero. It is clear that $\Ker\,\epsilon=\{\cA^{1}-\varepsilon\omega\cA^{1}\}$ 
is a subalgebra of $\C_{n+1}$. Moreover, the kernel of  
$\epsilon$ is a bilateral ideal of $\C_{n+1}$. Therefore, the algebra
$\C_{n}$, which we obtain in the result of the mapping $\epsilon:\;\C_{n+1}
\longrightarrow\C_{n}$, is {\it a quotient algebra}
\[ 
{}^\epsilon\C_{n}\simeq\C_{n+1}/\Ker\epsilon.
\]
Further, since the algebra $\C_{n}$ ($n=2m$) is isomorphic to the
full matrix
algebra $\M_{2^{m}}(\C)$, then in virtue of $\epsilon:\;\C_{n+1}
\longrightarrow\C_{n}\subset\C_{n+1}$ we obtain an homomorphic mapping
of $\C_{n+1}$ onto
the matrix algebra $\M_{2^{m}}(\C)$. 

The homomorphism $\epsilon:\;\cl_{p,q}\rightarrow\M_{2^m}(\R)$ is
analogously proved. In this case a quotient algebra has a form
\[
{}^\epsilon\cl_{p,q}\simeq\cl_{p+1,q}/\Ker\epsilon
\]
or
\[
{}^\epsilon\cl_{p,q}\simeq\cl_{p,q+1}/\Ker\epsilon,
\]
where $\Ker\epsilon=\{\cA^1-\omega\cA^1\}$, since in accordance with
(\ref{e3}) at $p-q\equiv 3,7\pmod{8}$ we have $\omega^2=1$, therefore
$\varepsilon=1$.
\end{proof}
 
Let us consider the form which the fundamental automorphisms of $\C_{n+1}$
take after the homomorphic mapping
$\epsilon:\;\C_{n+1}\rightarrow\C_{n}\subset\C_{n+1}$. First of all,
for the antiautomorphism $\cA\rightarrow\widetilde{\cA}$ it is necessary that
elements $\cA,\,\cB,\,\ldots\,\in\C_{n+1}$, which are mapped into one 
and the same element
$\cD\in\C_{n}$ (a kernel of the homomorphism $\epsilon$ if $\cD=0$) after the
transformation $\cA\rightarrow\widetilde{\cA}$ are must converted to the
elements $\widetilde{\cA},\,\widetilde{\cB},\,\ldots\,\in\C_{n+1}$, 
which are also
mapped into one and the same element $\widetilde{\cD}\in\C_{n}$. Otherwise, the
transformation $\cA\rightarrow\widetilde{\cA}$ is not transferred from
$\C_{n+1}$ into $\C_n$ as an unambiguous transformation. In particular, 
it is necessary in order that $\widetilde{\varepsilon\omega}=\varepsilon
\omega$, since $1$ and element $\varepsilon\omega$ under action of the
homomorphism $\epsilon$ are equally mapped into the unit, then $\widetilde{1}$
and $\widetilde{\varepsilon\omega}$ are also must be mapped into one 
and the same element in $\C_n$, but
$\widetilde{1}\rightarrow 1$, and $\widetilde{\varepsilon\omega}
\rightarrow\pm 1$ (in virtue of the formula (\ref{e8})). 
Therefore we must assume
\begin{equation}\label{e25}
\widetilde{\varepsilon\omega}=\varepsilon\omega.
\end{equation}
The condition (\ref{e25}) is sufficient for the transfer of the antiautomorphism
$\cA\rightarrow\widetilde{\cA}$ from $\C_{n+1}$ into $\C_{n}$. Indeed,
in this case we have
\[
\cA^{1}-\cA^{1}\varepsilon\omega\;\longrightarrow\;\widetilde{\cA^{1}}-
\widetilde{\varepsilon\omega}\widetilde{\cA^{1}}=\widetilde{\cA^{1}}-
\varepsilon\omega\widetilde{\cA^{1}}.
\]
Therefore, the elements of the form $\cA^{1}-\cA^{1}\varepsilon\omega$
(composing, as known, the kernel of $\epsilon$) under action of the 
transformation
$\cA\rightarrow\widetilde{\cA}$ are converted to the elements of the same
form.

The analogous conditions take place for other fundamental automorphisms.
However, for the automorphism $\cA\rightarrow\cA^{\star}$ a condition
$(\varepsilon\omega)^{\star}=\varepsilon\omega$ is not valid, since
$\omega$ is odd and in accordance with (\ref{e5}) we have
\begin{equation}\label{e26}
\omega^{\star}=-\omega.
\end{equation}
Thus, the automorphism $\cA\rightarrow\cA^{\star}$ is not transferred
from $\C_{n+1}$ into $\C_{n}$. 

Let us return to the antiautomorphism $\cA\rightarrow\widetilde{\cA}$ and
let consider in more details necessary conditions for the transfer of this
transformation from $\C_{n+1}$ to $\C_n$. First of all, the factor
$\varepsilon$ depending upon the condition $(\varepsilon\omega)^2=1$ and
the square of the element $\omega=\e_{12\ldots n+1}$ takes the following
values
\begin{equation}\label{e27}
\varepsilon=
\begin{cases}
1 & \text{if $p-q\equiv 3,7\pmod{8}$},\\
i & \text{if $p-q\equiv 1,5\pmod{8}$}.
\end{cases}
\end{equation}
Further, in accordance with (\ref{e8}) for the transformation $\omega
\rightarrow\widetilde{\omega}$ we obtain
\begin{equation}\label{e28}
\widetilde{\omega}=
\begin{cases}
\phantom{-}\omega & \text{if $p-q\equiv 3,7\pmod{8}$},\\
-\omega & \text{if $p-q\equiv 1,5\pmod{8}$}.
\end{cases}
\end{equation}
Therefore, for the algebras over the field $\F=\R$ the antiautomorphism
$\cA\rightarrow\widetilde{\cA}$ is transfered at the mappings $\cl_{p,q}
\rightarrow\cl_{p-1,q},\,\cl_{p,q}\rightarrow\cl_{p,q-1}$, where
$p-q\equiv 3,7\pmod{8}$. Over the field $\F=\C$ the antiautomorphism
$\cA\rightarrow\widetilde{\cA}$ is transfered in any case, since the
algebras $\C_{n+1}$ with signatures $p-q\equiv 3,7\pmod{8}$ and
$p-q\equiv 1,5\pmod{8}$ are isomorphic. In so doing, the condition (\ref{e25})
takes a form
\begin{eqnarray}
\widetilde{\omega}&=&\omega\quad\phantom{i}\text{if $p-q\equiv 3,7\pmod{8}$},
\nonumber\\
\widetilde{i\omega}&=&i\omega\quad\text{if $p-q\equiv 1,5\pmod{8}$}.\nonumber
\end{eqnarray}
Besides, each of these equalities satisfies the condition $(\varepsilon
\omega)^2=1$.

Let us consider now the antiautomorphism 
$\cA\rightarrow\widetilde{\cA^{\star}}$.
It is obvious that for the transfer of 
$\cA\rightarrow\widetilde{\cA^{\star}}$ from
$\C_{n+1}$ to $\C_{n}$ it is necessary that
\begin{equation}\label{e29}
\widetilde{(\varepsilon\omega)^{\star}}=\varepsilon\omega.
\end{equation}
It is easy to see that the mapping $\C_{p+q}\rightarrow\C_{p+q-1}$, where
$p+q\equiv 1,5\pmod{8}$, in virtue of (\ref{e26}) and the second equality
of (\ref{e28}), satisfies the condition (\ref{e29}), since in this case
\[
\widetilde{(\varepsilon\omega)^{\star}}=\varepsilon\widetilde{\omega^\star}=
-\varepsilon\omega^{\star}=\varepsilon\omega.
\]
Hence it immediately follows that over the field $\F=\R$ the antiautomorphism
$\cA\rightarrow\widetilde{\cA^\star}$ at the mappings $\cl_{p,q}\rightarrow
\cl_{p-1,q},\,\cl_{p,q}\rightarrow\cl_{p,q-1}$ ($p-q\equiv 3,7\pmod{8}$)
is not transferred.

Summarizing obtained above results we come to the following
\begin{theorem}\label{t8}
1) If $\F=\C$ and $\C_{p+q}\simeq\C_{p+q-1}\oplus\C_{p+q-1}$, where
$p+q\equiv 1,3,5,7\pmod{8}$, then the antiautomorphism $\cA\rightarrow
\widetilde{\cA}$ at the homomorphic mapping $\epsilon:\,\C_{p+q}\rightarrow
\C_{p+q-1}$ is transferred into a quotient algebra ${}^\epsilon\C_{p+q-1}$
in any case, the automorphism $\cA\rightarrow\cA^\star$ is not transferred,
and the antiautomorphism $\cA\rightarrow\widetilde{\cA^\star}$ is transferred
in the case of $p+q\equiv 1,5\pmod{8}$.\\
2) If $\F=\R$ and $\cl_{p,q}\simeq\cl_{p-1,q}\oplus\cl_{p-1,q},\,\cl_{p,q}
\simeq\cl_{p,q-1}\oplus\cl_{p,q-1}$, where $p-q\equiv 3,7\pmod{8}$, then
at the homomorphic mappings $\epsilon:\,\cl_{p,q}\rightarrow\cl_{p-1,q}$
and $\epsilon:\,\cl_{p,q}\rightarrow\cl_{p,q-1}$ the antiautomorphism
$\cA\rightarrow\widetilde{\cA}$ is transferred correspondingly into quotient
algebras ${}^\epsilon\cl_{p-1,q}$ and ${}^\epsilon\cl_{p,q-1}$ in any case,
and the automorphism $\cA\rightarrow\cA^\star$ and antiautomorphism
$\cA\rightarrow\widetilde{\cA^\star}$ are not transferred.
\end{theorem}
\section{Automorphism Groups of $\cl_{p,q}$, $\C_{p+q}$ and Discrete
Transformations of $O(p,q)$, $O(p+q,\C)$}
As noted above, there exists a close relation between D\c{a}browski groups
$\pin^{a,b,c}(p,q)$ and discrete tansformations of the orthogonal group
$O(p,q)$ (in particular, Lorentz group $O(1,3)$) \cite{DWGK,Ch97,Ch94,AlCh94,
AlCh96}. On the other hand, discrete transformations of the group $O(p,q)$
which acting in the space $\R^{p,q}$ associated with the algebra $\cl_{p,q}$,
may be realized via the fundamental automorphisms of $\cl_{p,q}$. In essence,
the group $\pin(p,q)$ is an intrinsic notion of $\cl_{p,q}$, since
in accordance with (\ref{e10}) $\pin(p,q)\subset\cl_{p,q}$. Let us show
that the D\c{a}browski group $\pin^{a,b,c}(p,q)$ is also completely defined
in the framework of the algebra $\cl_{p,q}$, that is, there is an equivalence
between $\pin^{a,b,c}(p,q)$ and the group $\pin(p,q)\subset\cl_{p,q}$
complemented by the transformations $\cA\rightarrow\cA^\star,\,\cA\rightarrow
\widetilde{\cA},\,\cA\rightarrow\widetilde{\cA^\star}$ (in connection with
this it should be noted that the Gauss-Klein group $\dZ_2\otimes\dZ_2$
is a finite group
corresponded to the algebra $\cl_{1,0}=\Om$ \cite{Sal81a,Sal84}).
\begin{prop}\label{p1}
Let $\cl_{p,q}$ ($p+q=2m$) be a Clifford algebra over the field $\F=\R$ and
let $\pin(p,q)$ be a double covering of the orthogonal group $O(p,q)=O_0(p,q)
\odot\{1,P,T,PT\}\simeq O_0(p,q)\odot(\dZ_2\otimes\dZ_2)$ of transformations
of the space $\R^{p,q}$, where $\{1,P,T,PT\}\simeq\dZ_2\otimes\dZ_2$ is a
group of discrete transformations of $\R^{p,q}$, $\dZ_2\otimes\dZ_2$ is the
Gauss-Klein group. Then there is an isomorphism between the group
$\{1,P,T,PT\}$ and an automorphism group $\{\Id,\star,\widetilde{\phantom{cc}},
\widetilde{\star}\}$ of the algebra $\cl_{p,q}$. In this case, parity
reversal $P$, time reversal $T$ and combination $PT$ are correspond 
respectively to the fundamental automorphisms $\cA\rightarrow\cA^\star,\,
\cA\rightarrow\widetilde{\cA}$ and $\cA\rightarrow\widetilde{\cA^\star}$.
\end{prop} 
\begin{proof} As known, the transformations $1,P,T,PT$ at the conditions
$P^2=T^2=(PT)^2=1,\;PT=TP$ form an abelian group with the following
multiplication table
\begin{center}{\renewcommand{\arraystretch}{1.4}
\begin{tabular}{|c||c|c|c|c|}\hline
    & $1$ & $P$ & $T$ & $PT$\\ \hline\hline
$1$ & $1$ & $P$ & $T$ & $PT$\\ \hline
$P$ & $P$ & $1$ & $PT$& $T$\\ \hline
$T$ & $T$ & $PT$& $1$ & $P$\\ \hline
$PT$& $PT$& $T$ & $P$ & $1$\\ \hline
\end{tabular}
}
\end{center}
Analogously, for the automorphism group $\{\Id,\star,\widetilde{\phantom{cc}},
\widetilde{\star}\}$ in virtue of the commutativity $\widetilde{(\cA^\star)}=
(\widetilde{\cA})^\star$ and the conditions $(\star)^2=(\widetilde{\phantom{cc}
})^2=\Id$ a following multiplication table takes place
\begin{center}{\renewcommand{\arraystretch}{1.4}
\begin{tabular}{|c||c|c|c|c|}\hline
        & $\Id$ & $\star$ & $\widetilde{\phantom{cc}}$ & $\widetilde{\star}$\\ \hline\hline
$\Id$   & $\Id$ & $\star$ & $\widetilde{\phantom{cc}}$ & $\widetilde{\star}$\\ \hline
$\star$ & $\star$ & $\Id$ & $\widetilde{\star}$ & $\widetilde{\phantom{cc}}$\\ \hline
$\widetilde{\phantom{cc}}$ & $\widetilde{\phantom{cc}}$ &$\widetilde{\star}$
& $\Id$ & $\star$ \\ \hline
$\widetilde{\star}$ & $\widetilde{\star}$ & $\widetilde{\phantom{cc}}$ &
$\star$ & $\Id$\\ \hline
\end{tabular}
}
\end{center}
The identity of the multiplication tables proves the isomorphism of the
groups $\{1,P,T,PT\}$ and $\{\Id,\star,\widetilde{\phantom{cc}},\widetilde{
\star}\}$.
\end{proof}
Further, in the case of anticommutativity $PT=-TP$ and $P^2=T^2=(PT)^2=\pm 1$
an isomorphism between the group $\{1,P,T,PT\}$ and an automorphism group
$\{\sI,\sW,\sE,\sC\}$, where $\sW,\sE$ and $\sC$ in accordance with (\ref{e17}),
(\ref{e15}) and (\ref{e19}) are the matrix representations of the
automorphisms $\cA\rightarrow\cA^\star,\,\cA\rightarrow\widetilde{\cA}$
and $\cA\rightarrow\widetilde{\cA^\star}$, is analogously proved.\\[0.4cm]
{\bf Example}. According to (\ref{e20}), (\ref{e21}), (\ref{e22}) and
(\ref{e23}) the matrix representation of the fundamental automorphisms
of the Dirac algebra $\C_4$ is defined by the following expressions:
$\sW=\gamma_1\gamma_2\gamma_3\gamma_4,\,\sE=\gamma_1\gamma_3,\,
\sC=\gamma_2\gamma_4$.
The multiplication table of the group $\{\sI,\sW,\sE,\sC\}\sim
\{I,\gamma_1\gamma_2
\gamma_3\gamma_4,\gamma_1\gamma_3,\gamma_2\gamma_4\}$ has a form 
\begin{multline}\label{e30}{\renewcommand{\arraystretch}{1.4}
\begin{tabular}{|c||c|c|c|c|}\hline
    & $I$ & $\gamma_1\gamma_2\gamma_3\gamma_4$ & $\gamma_1\gamma_3$ &
$\gamma_2\gamma_4$ \\ \hline\hline
$I$ & $I$ & $\gamma_1\gamma_2\gamma_3\gamma_4$ & $\gamma_1\gamma_3$ &
$\gamma_2\gamma_4$ \\ \hline
$\gamma_1\gamma_2\gamma_3\gamma_4$ & $\gamma_1\gamma_2\gamma_3\gamma_4$ &
$I$ & $\gamma_2\gamma_4$ & $\gamma_1\gamma_3$\\ \hline
$\gamma_1\gamma_3$ & $\gamma_1\gamma_3$ & $\gamma_2\gamma_4$ & $-I$ &
$-\gamma_1\gamma_2\gamma_3\gamma_4$\\ \hline
$\gamma_2\gamma_4$ & $\gamma_2\gamma_4$ & $\gamma_1\gamma_3$ & 
$-\gamma_1\gamma_2\gamma_3\gamma_4$ & $-I$ \\ \hline
\end{tabular}
}
\;\;\sim\;\;\\{\renewcommand{\arraystretch}{1.4}
\begin{tabular}{|c||c|c|c|c|}\hline  
    & $\sI$ & $\sW$ & $\sE$ & $\sC$\\ \hline\hline
$\sI$ & $\sI$ & $\sW$ & $\sE$ & $\sC$\\ \hline
$\sW$ & $\sW$ & $\sI$ & $\sC$ & $\sE$\\ \hline
$\sE$ & $\sE$ & $\sC$ & $-\sI$& $-\sW$\\ \hline
$\sC$ & $\sC$ & $\sE$ & $-\sW$& $-\sI$\\ \hline
\end{tabular}.
}
\end{multline}
However, in this representation we cannot directly to identify 
$\sW=\gamma_1\gamma_2\gamma_3\gamma_4$ with the parity reversal $P$, since
in this case the Dirac equation $(i\gamma_4\frac{\partial}{\partial x_4}-
i\boldsymbol{\gamma}\frac{\partial}{\partial\bx}-m)\psi(x_4,\bx)=0$ to be not
invariant with respect to $P$. On the other hand, for the canonical basis
(\ref{e20}) there exists a standard representation $P=\gamma_4,\,T=\gamma_1
\gamma_3$ \cite{BLP89}. The multiplication table of a group $\{1,P,T,PT\}\sim
\{I,\gamma_4,\gamma_2\gamma_3,\gamma_4\gamma_1\gamma_3\}$ has a form 
\begin{equation}\label{e31}{\renewcommand{\arraystretch}{1.4}
\begin{tabular}{|c||c|c|c|c|}\hline
   & $I$ & $\gamma_4$ & $\gamma_1\gamma_3$ & $\gamma_4\gamma_1\gamma_3$ 
\\ \hline\hline
$1$ & $I$& $\gamma_4$ & $\gamma_1\gamma_3$ & $\gamma_4\gamma_1\gamma_3$ 
\\ \hline
$\gamma_4$ & $\gamma_4$ & $I$ & $\gamma_4\gamma_1\gamma_3$ & $\gamma_1\gamma_3$
\\ \hline
$\gamma_1\gamma_3$ & $\gamma_1\gamma_3$ & $\gamma_4\gamma_1\gamma_3$ &
$-I$ & $-\gamma_4$ \\ \hline
$\gamma_4\gamma_1\gamma_3$ & $\gamma_4\gamma_1\gamma_3$ & $\gamma_1\gamma_3$
& $-\gamma_4$ & $-I$ \\ \hline
\end{tabular}
}
\;\;\sim\;\;{\renewcommand{\arraystretch}{1.4}
\begin{tabular}{|c||c|c|c|c|}\hline
    & $1$ & $P$ & $T$ & $PT$\\ \hline\hline
$1$ & $1$ & $P$ & $T$ & $PT$\\ \hline
$P$ & $P$ & $1$ & $PT$& $T$\\ \hline
$T$ & $T$ & $PT$& $-1$& $-P$\\ \hline
$PT$& $PT$& $T$ & $-P$& $-1$\\ \hline
\end{tabular}.
}
\end{equation}
It is easy to see that the tables (\ref{e30}) and (\ref{e31}) are equivalent,
therefore we have an isomorphism $\{\sI,\sW,\sE,\sC\}\simeq\{1,P,T,PT\}$. Besides,
each of these groups is isomorphic to the group $\dZ_4$.
\begin{theorem}\label{t9}
Let $\bsA=\{\sI,\,\sW,\,\sE,\,\sC\}$ be the automorphism group of the algebras  
$\cl_{p,q},\;
\C_{p+q}$ $(p+q=2m)$, where 
$\sW=\cE_1\cE_2\cdots\cE_m\cE_{m+1}\cE_{m+2}\cdots\cE_{p+q}$,
and $\sE=\cE_1\cE_2\cdots\cE_m$, $\sC=\cE_{m+1}\cE_{m+2}\cdots\cE_{p+q}$ if
$m\equiv 1\pmod{2}$, and $\sE=\cE_{m+1}\cE_{m+2}\cdots\cE_{p+q}$, 
$\sC=\cE_1\cE_2\cdots
\cE_m$ if $m\equiv 0\pmod{2}$. Let $\bsA_-$ and $\bsA_+$ be the automorphism 
groups, in which the all elements respectively commute
$(m\equiv 0\pmod{2})$ and anticommute $(m\equiv 1\pmod{2})$.
Then there are the following isomorphisms between finite groups and
automorphism groups with different signatures 
$(a,\,b,\,c)$, where $a,b,c\in\{-,+\}$:\\
1) $\F=\R$. $\bA_-\simeq\dZ_2\otimes\dZ_2$ for the signature $(+,\,+,\,+)$
if $p-q\equiv 0,4\pmod{8}$. $\bA_-\simeq\dZ_4$ for $(+,\,-,\,-)$ if
$p-q\equiv 0,4\pmod{8}$ and for $(-,\,+,\,-),\;(-,\,-,\,+)$ if
$p-q\equiv 2,6\pmod{8}$. $\bA_+\simeq Q_4/\dZ_2$ for $(-,\,-,\,-)$ if
$p-q\equiv 2,6\pmod{8}$. $\bA_+\simeq D_4/\dZ_2$ for $(-,\,+,\,+)$ if
$p-q\equiv 2,6\pmod{8}$ and for $(+,\,-,\,+),\,(+,\,+,\,-)$ if
$p-q\equiv 0,4\pmod{8}$.\\
2) Over the field $\F=\C$ there are only two non-isomorphic groups:
$\bA_-\simeq\dZ_2\otimes\dZ_2$ for the signature $(+,\,+,\,+)$ if
$p-q\equiv 0,4\pmod{8}$ and
$\bA_+\simeq Q_4/\dZ_2$ for 
 $(-,\,-,\,-)$ if
$p-q\equiv 2,6\pmod{8}$.
\end{theorem}
\begin{proof} First of all, since $\omega^2=+1$ if
$p-q\equiv 0,4\pmod{8}$ and $\omega^2=-1$ if $p-q\equiv 2,6\pmod{8}$,
then in the case of $\F=\R$ for the matrix of the automorphism 
$\cA\rightarrow\cA^\star$ we have
\[
\sW=\begin{cases}
+\sI, & \text{if $p-q\equiv 0,4\pmod{8}$};\\
-\sI, & \text{if $p-q\equiv 2,6\pmod{8}$}.
\end{cases}
\]
Over the field $\F=\C$ we can always to suppose $\sW^2=1$. Further, let us
find now the matrix $\sE$ of the antiautomorphism $\cA\rightarrow
\widetilde{\cA}$ at any $n=2m$, and elucidate the conditions at which the
matrix $\sE$ commutes with $\sW$, and also define a square of the matrix
$\sE$. Follows to \cite{Rash} let introduce along with the algebra $\C_{p+q}$
an auxiliary algebra $\C_m$ with basis elements
\[
1,\;\varepsilon_\alpha,\;\varepsilon_{\alpha_1\alpha_2}\;(\alpha_1<\alpha_2),\;
\varepsilon_{\alpha_1\alpha_2\alpha_3}\;(\alpha_1<\alpha_2<\alpha_3),\;
\ldots\;\varepsilon_{12\ldots m}.
\]
In so doing, linear operators $\hat{\cE}_i$ acting in the space
$\C^m$ associated with the algebra $\C_m$, are defined by a following rule
\begin{eqnarray}
&&\hat{\cE}_j\phantom{m+}\;:\;\Lambda\longrightarrow\Lambda\varepsilon_j,\nonumber\\
&&\hat{\cE}_{m+j}\;:\;\Lambda^1\longrightarrow-i\varepsilon_j\Lambda^1,\;
\Lambda^0\longrightarrow i\varepsilon_j\Lambda^0,\label{e32}
\end{eqnarray}
where $\Lambda$ is a general element of the auxiliary algebra $\C_m$,
 $\Lambda^1$ and
$\Lambda^0$ are correspondingly odd and even parts of $\Lambda$,
$\varepsilon_j$ are units of the auxiliary algebra, $j=1,2,\ldots m$. 
Analogously,
in the case of matrix representations of $\cl_{p,q}$ we have
\begin{eqnarray}
&&\hat{\cE}_j\phantom{m+}\;:\;\Lambda\longrightarrow\Lambda\beta_j\varepsilon_j,
\nonumber \\
&&\hat{\cE}_{m+j}\;:\;\Lambda^1\longrightarrow-\varepsilon_j\beta_{m+j}
\Lambda^1,\;
\Lambda^0\longrightarrow\varepsilon_j\beta_{m+j}\Lambda^0,\label{e33}
\end{eqnarray}
where $\beta_i$ are arbitrary complex numbers.
It is easy to verify that transposition of the matrices of so defined
operators gives
\begin{equation}\label{e34}
\cE^T_j=\cE_j,\quad \cE^T_{m+j}=-\cE_{m+j}.
\end{equation}
Further, for the antiautomorphism $\cA\rightarrow\widetilde{\cA}:\;
\widetilde{\sA}=
\sE\sA^T\sE^{-1}$, since in this case $\e_i\rightarrow\e_i$, it is sufficient
to select the matrix $\sE$ so that
\[
\sE\cE^T_i\sE^{-1}=\cE_i
\]
or taking into account (\ref{e34})
\begin{equation}\label{e35}
\sE\cE^T_j\sE^{-1}=\cE_j,\quad \sE\cE^T_{m+j}\sE^{-1}=-\cE_{m+j}. 
\end{equation}
Therefore, if $m$ is odd, then the matrix $\sE$ has a form
\begin{equation}\label{e36}
\sE=\cE_1\cE_2\ldots \cE_m,
\end{equation}
since in this case a product $\cE_1\cE_2\ldots \cE_m$ commutes
with all elements $\cE_j\;(j=1,\ldots m)$ and anticommutes with all
elements $\cE_{m+j}$.
Analogously, if $m$ is even, then
\begin{equation}\label{e37}
\sE=\cE_{m+1}\cE_{m+2}\ldots \cE_{p+q}.
\end{equation}
As required according to (\ref{e35}) in this case a product (\ref{e37}) 
commutes with $\cE_j$ and anticommutes with
$\cE_{m+j}$.

Let us consider now the conditions at which the matrix $\sE$ commutes or
anticommutes with $\sW$. Let $\sE=\cE_1\cE_2\ldots \cE_m$, where $m$ is odd, 
since
$\sW=\cE_1\ldots \cE_m\cE_{m+1}\ldots \cE_{p+q}$, then
\begin{eqnarray}
\cE_1\ldots \cE_m\cE_1\ldots \cE_m\cE_{m+1}\ldots \cE_{p+q}
&=&(-1)^{\frac{m(m-1)}{2}}
\sigma_1\sigma_2\ldots\sigma_m\cE_{m+1}\ldots \cE_{p+q},\nonumber\\
\cE_1\ldots \cE_m\cE_{m+1}\ldots \cE_{p+q}\cE_1\ldots \cE_m&=&
(-1)^{\frac{m(3m-1)}{2}}
\sigma_1\sigma_2\ldots\sigma_m\cE_{m+1}\ldots \cE_{p+q},\nonumber
\end{eqnarray}
where $\sigma_i$ are the functions of the form (\ref{e2}). It is easy to see
that in this case the elements 
$\sW$ and $\sE$ are always anticommute. Indeed, a comparison
$\frac{m(3m-1)}{2}\equiv\frac{m(m-1)}{2}\pmod{2}$ is equivalent to
$m^2\equiv 0,1\pmod{2}$, and since $m$ is odd, then we have always
$m^2\equiv 1\pmod{2}$. At $m$ is even and $\sE=\cE_{m+1}\ldots 
\cE_{p+q}$ it is easy to see that the matrices $\sW$ and $\sE$ 
are always commute $(m\equiv 0\pmod{2})$.
Further, let $r$ be a quantity of the elements $\cE_j$ of the product (\ref{e36})
whose squares equal to $+\sI$, and let $s$ be a quantity of the elements
$\cE_{m+j}$ of the product (\ref{e36}) whose squares equal to $-\sI$. 
Then a square
of the matrix (\ref{e36}) 
at $m$ is odd  equals to $+\sI$ if $r-s\equiv 3\pmod{4}$ and respectively
$-\sI$ if $r-s\equiv 1\pmod{4}$. Analogously, a square of the matrix (\ref{e37})
at $m$ is even equals to $+\sI$ if $k-t\equiv
0\pmod{4}$ and respectively $-\sI$ if $k-t\equiv 2\pmod{4}$. It is obvious that
over the field $\F=\C$ we can to suppose $\cE^2_{1\ldots m}=\cE^2_{m+1\ldots
p+q}=\sI$.

Let us find now the matrix $\sC$ of the antiautomorphism $\cA\rightarrow
\widetilde{\cA^\star}$: $\widetilde{\sA^\star}=\sC\sA^T\sC^{-1}$. Since
in this case $\e_i\rightarrow-\e_i$, then it is sufficient to select the
matrix $\sC$ so that
\[
\sC\cE^T_i\sC^{-1}=-\cE_i
\]
or taking into account (\ref{e34})
\begin{equation}\label{e38}
\sC\cE^T_j\sC^{-1}=-\cE_j,\quad \sC\cE^T_{m+j}\sC^{-1}=\cE_{m+j}.
\end{equation}
where $j=1,\ldots m$.
In comparison with (\ref{e35}) it is easy to see that in
(\ref{e38}) the matrices $\cE_j$ and $\cE_{m+j}$ are changed by the roles.
Therefore, if $m$ is odd, then
\begin{equation}\label{e39}
\sC=\cE_{m+1}\cE_{m+2}\ldots \cE_{p+q},
\end{equation}
and if $m$ is even, then
\begin{equation}\label{e40}
\sC=\cE_1\cE_2\ldots \cE_m.
\end{equation}  
Permutation conditions of the matrices $\sC$ and $\sW$ are analogous to the
permutation conditions of $\sE$ with $\sW$, that is, the matrix $\sC$ of the form
(\ref{e39})
always anticommutes with $\sW$ ($m\equiv 1\pmod{2}$), and the matrix $\sC$ of the
form
(\ref{e40}) always commutes with $\sW$ ($m\equiv 0\pmod{2}$). Correspondingly,
a square of the matrix (\ref{e39}) equals to $+\sI$ if $k-t\equiv 3\pmod{4}$ and
$-\sI$ if $k-t\equiv 1\pmod{4}$. Analogously, a square of the matrix (\ref{e40})
equals to $+\sI$ if $r-s\equiv 0\pmod{4}$ and $-\sI$ if $r-s\equiv 2\pmod{4}$.
Obviously, over the field $\F=\C$ we can suppose $\sC^2=\sI$.

Finally, let us find permutation conditions of the matrices $\sE$ and $\sC$.
First of all, at $m$ is odd $\sE=\cE_1\ldots \cE_m$,
$\sC=\cE_{m+1}\ldots \cE_{p+q}$, alternatively, at $m$ is even 
$\sE=\cE_{m+1}\ldots
\cE_{p+q}$, $\sC=\cE_1\ldots \cE_m$. Therefore,
\[
\cE_1\ldots \cE_m\cE_{m+1}\ldots \cE_{p+q}=(-1)^{m^2}\cE_{m+1}\ldots \cE_{p+q}
\cE_1\ldots
\cE_m,
\]
that is, the matrices $\sE$ and $\sC$ commute at $m\equiv 0\pmod{2}$ and anticommute
at $m\equiv 1\pmod{2}$.

Now we have all the necessary conditions for the definition and
classification of isomorphisms between finite groups and automorphism groups
of Clifford algebras. Let $\F=\R$ and let $m\equiv 0\pmod{2}$, therefore,
a group $\bsA=\{\sI,\,\sW,\,\sE,\,\sC\}$ is Abelian. A condition 
$\sW^2=\sE^2=\sC^2=\sI$ 
is equivalent to $p-q\equiv 0,4\pmod{8},\,r-s\equiv 0\pmod{4},
\,k-t\equiv 0\pmod{4}$,
which, clearly, are compatible. In accordance with (\ref{e36})--(\ref{e38})
and (\ref{e39})--(\ref{e40}) at $m\equiv 0\pmod{2}$ $\sW=\sC\sE$ and 
$\sW^2=(\sC\sE)^2=
\sC\sE\sC\sE=\sC\sC\sE\sE=\sI$. Therefore, 
$\bsA_-\simeq\dZ_2\otimes\dZ_2$ for the signature
$(+,\,+,\,+)$ if $p-q\equiv 0,4\pmod{8}$. 
The isomorphism $\bsA_-\simeq\dZ_4$ for the signatures $(+,\,-,\,-)$ 
($p-q\equiv 0,4\pmod{8}$) and $(-,\,+,\,-),\;(-,\,-,\,+)$ 
($p-q\equiv 2,6\pmod{8}$) is analogously proved. It is easy to see that
for $m\equiv 0\pmod{2}$
there are only four isomorphisms considered previously. Further, for
$m\equiv 1\pmod{2}$ all the elements of the group $\bsA$ anticommute and in
this case $\sW=\sE\sC$. The signature $(-,\,-,\,-)$ is equivalent to conditions
$p-q\equiv 2,6\pmod{8},\,r-s\equiv 1\pmod{4},\,k-t\equiv 1\pmod{4}$,
here $\sW^2=(\sE\sC)^2=\sE\sC\sE\sC=-\sE\sE\sC\sC=-\sI$ 
and we have an isomorphism $\bsA_+\simeq
Q_4/\dZ_2$, where $Q_4/\dZ_2=\{1,\,\bi,\,\bj,\,\bk\}$, $\bi,\,\bj,\,\bk$
are the quaternion units. It is easy to verify that for the signatures
$(-,\,+,\,+)$ at $p-q\equiv 2,6\pmod{8}$ and $(+,\,-,\,+),\;
(+,\,+,\,-)$ at $p-q\equiv 0,4\pmod{8}$ we have an isomorphism
$\bsA_+\simeq D_4/\dZ_2$, where $D_4/\dZ_2=\{1,\,\e_1,\,\e_2,\,\e_{12}\}$,
$\e_1,\e_2$ are the units of the algebra $\cl_{1,1}$ or $\cl_{2,0}$.
The eight automorphism groups considered previously, each of which is
isomorphic to one from the four finite groups
 $\dZ_2\otimes\dZ_2,\,\dZ_4,\,
Q_4/\dZ_2,\,D_4/\dZ_2$, are the only possible over the field 
$\F=\R$. In contrast with this, over the field $\F=\C$ we can suppose 
$\sW^2=\sE^2=\sC^2=\sI$. At $m\equiv 0\pmod{2}$ we have only one signature
$(+,\,+,\,+)$ and an isomorphism $\bsA_-\simeq
\dZ_2\otimes\dZ_2$ if $p+q\equiv 0,4\pmod{8}$, since over the field $\C$ the
signatures
$(+,\,-,\,-),\;(-,\,+,\,-)$ and $(-,\,-\,-)$ are isomorphic to $(+,\,+,\,+)$.
Correspondingly, at $m\equiv 1\pmod{2}$ we have an isomorphism
$\bsA_+\simeq Q_4/\dZ_2$ for the signature $(-,\,-,\,-)$ if
$p+q\equiv 2,6\pmod{8}$. It should be noted that the signatures
$(+,\,+,\,+)$ and
$(-,\,-,\,-)$ are non-isomorphic, since there exists no a group $\bsA$ with the
signature $(+,\,+,\,+)$ in which all the elements anticommute, and also
there exists no a group $\bsA$ with $(-,\,-,\,-)$ in which the all elements
commute. Thus, over the field $\C$ we have only two
non-isomorphic automorphism groups: $\bsA_-\simeq\dZ_2\otimes\dZ_2,\;
\bsA_+\simeq Q_4/\dZ_2$.
\end{proof}

The following Theorem is a direct consequence of the previous Theorem.
Here we establish a relation between signatures $(a,b,c)$ of the
D\c{a}browski groups and signatures $(p,q)$ of the Clifford algebras with
even dimensionality.
\begin{theorem}\label{t10}
Let $\pin^{a,b,c}(p,q)$ be a double covering of the orthogonal group
$O(p,q)$ of the space $\R^{p,q}$ associated with the algebra
$\cl_{p,q}$ and let $\pin^{a,b,c}(p+q,\C)$ be a double covering of the complex
orthogonal group $O(p+q,\C)$ of the space $\C^{p+q}$ associated with the
algebra $\C_{p+q}$. Dimensionalities of the algebras
$\cl_{p,q}$ and $\C_{p+q}$ are even $(p+q=2m)$, squares of the
symbols $a,b,c\in
\{-,+\}$ are correspond to squares of the elements of the finite group 
$\bsA=\{\sI,\sW,\sE,\sC\}:\;a=\sW^2,\,b=\sE^2,\,c=\sC^2$, where $\sW,\sE$
 and $\sC$
are correspondingly the matrices of the fundamental automorphisms $\cA\rightarrow
\cA^\star,\,\cA\rightarrow\widetilde{\cA}$ and $\cA\rightarrow
\widetilde{\cA^\star}$ of $\cl_{p,q}$ and $\C_{p+q}$. Then over the field
$\K=\R$ for the algebra $\cl_{p,q}$ there are eight double coverings
of the group $O(p,q)$ and two non-isomorphic double coverings of the group
$O(p+q,\C)$ for $\C_{p+q}$ over the field $\K=\C$:\\
1) $\F=\R$. Non--Cliffordian groups
\[
\pin^{+,+,+}(p,q)\simeq\frac{(\spin_0(p,q)\odot\dZ_2\otimes\dZ_2\otimes\dZ_2)}
{\dZ_2},
\]
if $p-q\equiv 0,4\pmod{8}$ and
\[
\pin^{a,b,c}(p,q)\simeq\frac{(\spin_0(p,q)\odot(\dZ_2\otimes\dZ_4)}{\dZ_2},
\]\begin{sloppypar}\noindent
if $(a,b,c)=(+,-,-)$ and $p-q\equiv 0,4\pmod{8}$, and also
$(a,b,c)=\{(-,+,-),\,(-,-,+)\}$ if $p-q\equiv 2,6\pmod{8}$.\\
\end{sloppypar}\noindent
Cliffordian groups
\[
\pin^{-,-,-}(p,q)\simeq\frac{(\spin_0(p,q)\odot Q_4)}{\dZ_2},
\]
if $p-q\equiv 2,6\pmod{8}$ and
\[
\pin^{a,b,c}(p,q)\simeq\frac{(\spin_0(p,q)\odot D_4)}{\dZ_2},
\]\begin{sloppypar}\noindent
if $(a,b,c)=(-,+,+)$ and $p-q\equiv 2,6\pmod{8}$, and also if
$(a,b,c)=\{(+,-,+),\,(+,+,-)\}$ and $p-q\equiv 0,4\pmod{8}$.\\
\end{sloppypar}
2) $\F=\C$. A non-Cliffordian group
\[
\pin^{+,+,+}(p+q,\C)\simeq
\frac{(\spin_0(p+q,\C)\odot\dZ_2\otimes\dZ_2\otimes\dZ_2)}
{\dZ_2},
\]
if $p+q\equiv 0,4\pmod{8}$. A Cliffordian group
\[
\pin^{-,-,-}(p+q,\C)\simeq\frac{(\spin_0(p+q,\C)\odot Q_4)}{\dZ_2},
\]
if $p+q\equiv 2,6\pmod{8}$.
\end{theorem}
\section{D\c{a}browski Groups for Odd-dimensional Spaces}
According to Theorem \ref{t8} and Theorem \ref{t4}
in the case of odd-dimensional spaces $\R^{p,q}$ and $\C^{p+q}$ 
the algebra homomorphisms $\cl_{p,q}\rightarrow
\cl_{p-1,q},\,\cl_{p,q}\rightarrow\cl_{p,q-1}$ and $\C_{p+q}\rightarrow
\C_{p+q-1}$ induce group homomorphisms $\pin(p,q)\rightarrow\pin(p-1,q),\,
\pin(p,q)\rightarrow\pin(p,q-1)$, $\pin(p+q,\C)\rightarrow\pin(p+q-1,\C)$
and correspondingly $\pin(p,q)\rightarrow\spin(p,q),\,\pin(p+q,\C)\rightarrow
\spin(p+q,\C)$.
\begin{theorem}\label{t11}
1) If $\F=\R$ and $\pin^{a,b,c}(p,q)\simeq\pin^{a,b,c}(p-1,q)\cup\omega
\pin^{a,b,c}(p-1,q),\,\pin^{a,b,c}(p,q)\simeq\pin^{a,b,c}(p,q-1)\cup\omega
\pin^{a,b,c}(p,q-1)$ are the D\c{a}browski groups over $\R$, where $p-q\equiv 3,7
\pmod{8}$, then in the result of homomorphic mappings $\pin^{a,b,c}(p,q)
\rightarrow\pin^{a,b,c}(p-1,q)$ and $\pin^{a,b,c}(p,q)\rightarrow
\pin^{a,b,c}(p,q-1)$ take place following quotient groups:
\begin{eqnarray}
\pin^b(p-1,q)&\simeq&\frac{(\spin_0(p-1,q)\odot\dZ_2\otimes\dZ_2)}{\dZ_2},
\nonumber\\
\pin^b(p,q-1)&\simeq&\frac{(\spin_0(p,q-1)\odot\dZ_2\otimes\dZ_2)}{\dZ_2}.
\nonumber
\end{eqnarray}
2) If $\F=\C$ and $\pin^{a,b,c}(p+q,\C)\simeq\pin^{a,b,c}(p+q-1,\C)\cup
\pin^{a,b,c}(p+q-1,\C)$ are the D\c{a}browski groups over $\C$, where $p+q\equiv
1,3,5,7\pmod{8}$, then in the result of an homomorpic mapping $\pin^{a,b,c}
(p+q,\C)\rightarrow\pin^{a,b,c}(p+q-1,\C)$ take place following quotient
groups:
\[
\pin^b(p+q-1,\C)\simeq\frac{(\spin_0(p+q-1,\C)\odot\dZ_2\otimes\dZ_2)}{\dZ_2},
\]
if $p+q\equiv 3,7\pmod{8}$ and
\[
\pin^{b,c}(p+q-1,\C),
\]
if $p+q\equiv 1,5\pmod{8}$, at this point a set of the fundamental automorphisms,
which correspond to the discrete transformations of the space
$\C^{p+q-1}$ associated with a quotient algebra
${}^\epsilon\C_{p+q-1}$, does not form a finite group.
\end{theorem}
\begin{proof} Indeed, over the field $\F=\R$ in accordance with Theorem
\ref{t8} from all the fundamental automorphisms
at the homomorphic mappings $\cl_{p,q}\rightarrow\cl_{p-1,q}$
and $\cl_{p,q}\rightarrow\cl_{p,q-1}$ only the antiautomorphism
$\cA\rightarrow\widetilde{\cA}$ is transferred into
quotient algebras ${}^\epsilon\cl_{p-1,q}$
and ${}^\epsilon\cl_{p,q-1}$. Further, according to Proposition
\ref{p1} the antiautomorphism $\cA\rightarrow\widetilde{\cA}$
corresponds to time reversal $T$. Therefore, groups of the discrete
transformations of the spaces $\R^{p-1,q}$ and $\R^{p,q-1}$
associated with the quotient algebras ${}^\epsilon\cl_{p-1,q}$ and
${}^\epsilon\cl_{p,q-1}$ are defined by a two--element group $\{1,T\}\sim
\{\sI,\sE\}\simeq\dZ_2$, where $\{\sI,\sE\}$ is an automorphism group of the quotient
algebras
${}^\epsilon\cl_{p-1,q},\,{}^\epsilon\cl_{p-1,q}$, $\sE$ is a matrix of the
antiautomorphism $\cA\rightarrow\widetilde{\cA}$. Thus, at
$p-q\equiv 3,7\pmod{8}$ there are the homomorphic mappings
$\pin^{a,b,c}(p,q)\rightarrow\pin^b(p-1,q)$ and $\pin^{a,b,c}(p,q)\rightarrow
\pin^b(p,q-1)$, where $\pin^b(p-1,q),\,\pin^b(p,q-1)$ are quotient groups,
$b=T^2=\sE^2$. At this point, a double covering of $C^b$ is isomorphic to
$\dZ_2\otimes\dZ_2$.

Analogously, over the field $\F=\C$ at $p+q\equiv 3,7\pmod{8}$ we have 
a quotient group $\pin^b(p+q-1,\C)$. Further, according to Theorem
\ref{t8} in the result of the homomorphic mapping $\C_{p+q}\rightarrow
\C_{p+q-1}$ the antiautomorphisms $\cA\rightarrow\widetilde{\cA},\,\cA
\rightarrow\widetilde{\cA^\star}$ are transferred into a quotient algebra
${}^\epsilon\C_{p+q-1}$ at $p+q\equiv 1,5\pmod{8}$.
Therefore, a set of the discrete transformations of the space 
$\C^{p+q-1}$ associated with the quotient algebra ${}^\epsilon\C_{p+q-1}$
is defined by a three-element set $\{1,T,PT\}\sim\{\sI,\sE,\sC\}$, where
$\{\sI,\sE,\sC\}$ is a set of the automorphisms of ${}^\epsilon\C_{p+q-1}$,
$\sE$ and $\sC$ are correspondingly the matrices of the antiautomorphisms
$\cA\rightarrow
\widetilde{\cA}$ and $\cA\rightarrow\widetilde{\cA^\star}$. It is easy to see
that the set $\{1,T,PT\}\sim\{\sI,\sE,\sC\}$ does not form a finite group. Thus,
at $p+q\equiv 1,5\pmod{8}$ there is an homomorphism $\pin^{a,b,c}(p+q,\C)
\rightarrow\pin^{b,c}(p+q-1,\C)$, where $\pin^{b,c}(p+q-1,\C)$ is a quotient
group, $b=T^2=\sE^2,\,c=(PT)^2=\sC^2$.
\end{proof}
{\bf Example}. Let us consider a simplest complex Clifford algebra with
odd dimensionality, $\C_3$. The algebra $\C_3$ may be represented by two
different complexifications:
$\C_3=\C\otimes\cl_{3,0}$
and $\C_3=\C\otimes\cl_{0,3}$, where $\cl_{3,0}$ and $\cl_{0,3}$ are
correspondingly the algebras of hyperbolic and elliptic biquaternions.
In accordance with Theorem \ref{t2} there is a decomposition of $\C_3$ 
into a direct sum of two subalgebras, which may be represented by a 
following scheme:
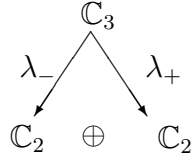
\[
\unitlength=0.5mm
\begin{picture}(70,50)
\put(35,40){\vector(2,-3){15}}
\put(35,40){\vector(-2,-3){15}}
\put(32.25,42){$\C_{3}$}
\put(16,28){$\lambda_{-}$}
\put(49.5,28){$\lambda_{+}$}
\put(13.5,9.20){$\C_{2}$}
\put(52.75,9){$\C_{2}$}
\put(32.5,10){$\oplus$}
\end{picture}
\]
Here the idempotents
\[
\lambda_{-}=\frac{1-i\omega}{2},\quad
\lambda_{+}=\frac{1+i\omega}{2} 
\]
in accordance with \cite{CF97}
may be identified with helicity projection operators. Further, according to
Theorem \ref{t8} at the homomorphic mapping $\epsilon:\,\C_3
\rightarrow\C_2$ the antiautomorphisms
$\cA\rightarrow\widetilde{\cA}$ and $\cA\rightarrow\widetilde{\cA^\star}$
are transferred into a quotient algebra ${}^\epsilon\C_2$, and the
automorphism $\cA\rightarrow\cA^\star$ is not transferred. Therefore,
there is an homomorphism $\pin^{a,b,c}(3,\C)\rightarrow\pin^{b,c}(2,\C)$
(Theorem \ref{t11}), where $\pin^{b,c}(2,\C)$ is a quotient group which
double covers the orthogonal group $O(2,\C)$ of the space
$\C^2$ associated with ${}^\epsilon\C_2$.
According to proposition \ref{p1} the automorphism $\cA\rightarrow\cA^\star$
corresponds to parity reversal $P$ which under action of the 
homomorphism $\epsilon$ is not transferred into the quotient algebra
${}^\epsilon\C_2$ and correspondingly quotient group $\pin^{b,c}(2,\C)$.
Thus, we have a `symmetry breaking' of the group of discrete transformations
with excluded operation $P$. In physics there is an analog of this
situation known as a {\it parity violation}. In connection with this it pays
to relate the quotient algebra ${}^\epsilon\C_2$ and quotient group
$\pin^{b,c}(2,\C)$ with some chiral field, for example, neitrino field.
As known, in Nature there exist only left neitrino and right antineitrino and
there exist no right neitrino and left antineitrino, therefore, for the
neitrino field the operation $P$ is violated. In order to proceed this
analogy, at first we must to establish a relation between the
quotient algebra ${}^\epsilon\C_2$ and some spinor field.
Let us show that such a field is a Dirac-Hestenes spinor field
\cite{Hest66,Hest90}. Indeed, in accordance with Theorem \ref{t1} we have
$\C_2\simeq\cl_{3,0}$, further $\cl_{3,0}\simeq\cl^+_{1,3}$, where $\cl_{1,3}$
is a spacetime algebra. Units of the algebra $\cl_{1,3}$ in the matrix
representation have the form
\begin{equation}\label{e41}\ar
\Gamma_0=\begin{pmatrix}
I & 0\\
0 & -I
\end{pmatrix},\;\;\Gamma_1=\begin{pmatrix}
0 & \sigma_1\\
-\sigma_1 & 0
\end{pmatrix},\;\;\Gamma_2=\begin{pmatrix}
0 & \sigma_2\\
-\sigma_2 & 0
\end{pmatrix},\;\;\Gamma_3=\begin{pmatrix}
0 & \sigma_3\\
-\sigma_3 & 0
\end{pmatrix},
\end{equation}
where $\sigma_i$ are the Pauli matrices
\[
\ar
\sigma_1=\begin{pmatrix}
0 & 1\\
1 & 0
\end{pmatrix},\quad\sigma_2=\begin{pmatrix}
0 & -i\\
i & 0
\end{pmatrix},\quad\sigma_3=\begin{pmatrix}
1 & 0\\
0 &-1
\end{pmatrix},
\] 
$I$ is the unit matrix. The Dirac-Hestenes spinor
$\phi$ is an element of the algebra $\cl^+_{1,3}\simeq\cl_{3,0}$
and, therefore, may be represented by the biquaternion number
\begin{equation}
\phi=a^0+a^{01}\Gamma_{01}+a^{02}\Gamma_{02}+a^{03}\Gamma_{03}+
a^{12}\Gamma_{12}+a^{13}\Gamma_{13}+a^{23}\Gamma_{23}+a^{0123}\Gamma_{0123}.
\label{e42}
\end{equation}
Or in the matrix form
\begin{equation}\label{e43}\ar
\phi=\begin{pmatrix}
\phi_1 & -\phi^\ast_2 & \phi_3 & \phi^\ast_4 \\
\phi_2 & \phi^\ast_1 & \phi_4 & -\phi^\ast_3\\
\phi_3 & \phi^\ast_4 & \phi_1 & -\phi^\ast_2\\
\phi_4 & -\phi^\ast_3 & \phi_2 & \phi^\ast_1
\end{pmatrix},
\end{equation}
where
\begin{eqnarray}
\phi_1&=&a^0-ia^{12},\nonumber\\
\phi_2&=&a^{13}-ia^{23},\nonumber\\
\phi_3&=&a^{03}-ia^{0123},\nonumber\\
\phi_4&=&a^{01}+ia^{02}.\nonumber
\end{eqnarray}
The spinor $\phi$, defined by the expression (\ref{e42}) or (\ref{e43}),
satisfies to a Dirac-Hestenes equation \cite{Hest90}
\begin{equation}\label{e44}
\partial\phi\Gamma_{21}=\frac{mc}{\hbar}\phi\Gamma_0,
\end{equation}
where $\partial=\Gamma^\nu\frac{\partial}{\partial x^\nu}$. Further, let
${}^\epsilon\phi\in{}^\epsilon\C_2$ be a Dirac-Hestenes 'quotient spinor'.
In virtue of the isomorphism ${}^\epsilon\C_2\simeq\cl_{3,0}\simeq\cl^+_{1,3}$ 
we have for the spinor ${}^\epsilon\phi$ the representation (\ref{e42}).
It is known \cite{RSVL} that a transformation group of the Dirac-Hestenes
field is a group $\spin_+(1,3)$ which double covers the connected component
of the Lorentz group, since
\[
\ar
\spin_+(1,3)\simeq\left\{\begin{pmatrix} a & b \\ c & d\end{pmatrix}\in\C_2:\;
\det\begin{pmatrix} a & b \\ c & d\end{pmatrix}=1\right\}=SL(2,\C).
\]
It is obvious that $\pin(2,\C)\simeq\spin(3,\C)$ and further $\spin(3,\C)\simeq
\spin(1,3)$, whence $\spin_0(3,\C)\simeq\spin_+(1,3)$, therefore,
a transformation group of the quotient spinor ${}^\epsilon\phi$ is also
isomorphic to $\spin_+(1,3)$. The complete transformation group of
${}^\epsilon\phi$ is isomorphic to the quotient group
$\pin^{b,c}(2,\C)$ consisting of 
$\spin_+(1,3)$ and the two discrete transformations $T$ and $PT$ (the latter
transformation in virtue of $CPT$-theorem is equivalent to a charge
conjugation $C$). Since the quotient algebra ${}^\epsilon\C_2$ is isomorphic
to a subalgebra of all the even elements of the spacetime algebra  $\cl_{1,3}$,
then we can express the automorphisms of ${}^\epsilon\C_2$ via the
automorphisms of $\cl_{1,3}$. In accordance with Theorem \ref{t9}
the automorphism group of $\cl_{1,3}$ is isomorphic to the Abelian group
$\dZ_4$. The non-Cliffordian group double covering the Lorentz group $O(1,3)$ 
has a form (theorem \ref{t10})  
\begin{eqnarray}
\pin^{-,-,+}(1,3)&\simeq&\frac{(\spin_0(1,3)\odot\dZ_2\otimes\dZ_4)}{\dZ_2}
\nonumber\\
&\simeq&\frac{(SL(2,\C)\odot\dZ_2\otimes\dZ_4)}{\dZ_2}.\nonumber
\end{eqnarray}
In the group $\bsA_-=\{\sI,\sW,\sE,\sC\}$ the matrices $\sE$ and $\sC$ of the
antiautomorphisms $\cA\rightarrow\widetilde{\cA}$ and 
$\cA\rightarrow\widetilde{\cA^\star}$
for the basis (\ref{e41}) have the form 
\[
\ar
\sE=\begin{pmatrix}
0 & 1 & 0 & 0\\
-1& 0 & 0 & 0\\
0 & 0 & 0 & 1\\
0 & 0 & -1& 0
\end{pmatrix},\quad \sC=\begin{pmatrix}
0 & 0 & 0 & -i\\
0 & 0 & i & 0\\
0 & -i& 0 & 0\\
i & 0 & 0 & 0
\end{pmatrix}
\]
Further, in accordance with (\ref{e19}) an action of the
antiautomorphism $\cA\rightarrow
\widetilde{\cA^\star}$ on the spinor $\phi$, defined by the matrix
representation (\ref{e43}), is expressed as follows
\begin{equation}\label{e45}\ar
\widetilde{\phi^\star}=\begin{pmatrix}
\phi^\ast_1 & \phi^\ast_2 & -\phi^\ast_3 & -\phi^\ast_4 \\
-\phi_2 & \phi_1 & -\phi_4 & \phi_3\\
-\phi^\ast_3 & -\phi^\ast_4 & \phi^\ast_1 & \phi^\ast_2\\
-\phi_4 & \phi_3 & -\phi_2 & \phi_1
\end{pmatrix}
\end{equation}
It should be noted that the same result may be obtained via $\Gamma^0
\phi^+\Gamma^0$ (see \cite{VR93}).
Let us assume now that a massless Dirac-Hestenes equation (\ref{e44})
describes the neitrino field
\begin{equation}\label{e46}
\partial{}^\epsilon\phi\Gamma_{21}=0,
\end{equation}
then the antineitrino field is described by an equation
\begin{equation}\label{e47}
\widetilde{\left(\partial{}^\epsilon\phi\Gamma_{21}\right)^\star}=0.
\end{equation}
The equations (\ref{e46}) and (\ref{e47}) are converted into each other under
action of the antiautomorphism $\cA\rightarrow\widetilde{\cA^\star}$ which
in accordance with Proposition \ref{p1} corresponds to the combination $PT$
(charge conjugation by $CPT$--theorem). The fields describing by the equations
(\ref{e46}) and (\ref{e47}) possess a fixed helicity (there exist no right 
neitrino and left antineitrino), since the automorphism
$\cA\rightarrow\cA^\star$ corresponded to parity reversal
$P$ in this case is not defined. On the other hand, in accordance with the
Feynman-Stueckelberg interpretation the antiparticles are considered as
particles moving back in time, therefore a time-reversed equation
(\ref{e46}) describes the antiparticle (antineitrino). Moreover, in the
Feynman--Stueckelberg interpretation time reversal for the chiral field
gives rise to the well--known $CP$--invariance in the theory of weak 
interactions. Thus, the actions of the antiautomorphisms $\cA\rightarrow
\widetilde{\cA}$ and $\cA\rightarrow\widetilde{\cA^\star}$ and the
corresponding operations $T$ and $PT\sim C$ on the field ${}^\epsilon\phi$
are equivalent. This equivalence immediately follows from (\ref{e8})
and (\ref{e9}) (see also \cite{FRO90b}), namely, for $\cA\in\cl^+_{p,q}$
we have always $\widetilde{\cA}=\widetilde{\cA^\star}$, in our case
${}^\epsilon\phi\in\cl^+_{1,3}$ and $\widetilde{{}^\epsilon\phi}=
\widetilde{{}^\epsilon\phi^\star}$.
\section*{Acknowledgments} I am grateful to Prof. D\c{a}browski for
sending me his interesting works.

\begin{thebibliography}{Clifford}
\bibitem[AlCh94]{AlCh94} L.J. Alty, A. Chamblin, {\it Spin Structures on
Kleinian Manifolds}, Class. Quantum Grav. {\bf 11}, 2411-2415, (1994).
\bibitem[AlCh96]{AlCh96} L.J. Alty, A. Chamblin, {\it Obstructions to Pin
Structures on Kleinian Manifolds}, J. Math. Phys. {\bf 37}, 2001-2011 , (1996).
\bibitem[Amm98]{Amm98} B. Ammann, {\bf Spin-Strukturen und das Spektrum des
Dirac-Operators} (Dissertation Freiburg 1998, Shaker-Verlag Aachen 1998). 
\bibitem[AtBSh]{AtBSh} M.F. Atiyah, R. Bott, A. Shapiro, {\it Clifford
modules}, Topology, {\bf 3}, (Suppl. 1),3-38, (1964).
\bibitem[B\"{a}r91]{Bar91} C. B\"{a}r, {\bf Das Spektrum von Dirac-Operatoren}
(Dissertation, Bonner Math. Schriften {\bf 217}, 1991).
\bibitem[Bau81]{Bau81} H. Baum, {\bf Spin-Strukturen und Dirac-Operatoren
\"{u}ber pseudoriemannschen Mannigfaltigkeiten} (Teubner, Leipzig, 1981).  
\bibitem[BLP89]{BLP89} V.B. Berestetskii, E.M. Lifshitz, L.P. Pitaevskii,
{\bf Quantum electrodynamics} (Nauka, Moscow 1989) [in Russian].
\bibitem[BD89]{BD89} M. Blau, L. D\c{a}browski, {\it Pin structures on manifolds
quotiented by discrete groups}, J. Geometry and Physics {\bf 6}, 143-157,
(1989).
\bibitem[BH]{BH} A. Borel and F. Hirzebruch, {\it Characteristic classes
and homogeneous spaces}, Amer. J. Math. {\bf 80}, 458-538, {\bf 81} 315-382
and {\bf 82} 491-504 (1958, 1959 and 1960).  
\bibitem[CGT95]{CGT95} M. Cahen, S. Gutt and A. Trautman, {\it Pin structures
and the modified Dirac operator}, J. Geometry and Physics {\bf 17}, 283-297
(1995).
\bibitem[Ch94a]{Ch94a} A. Chamblin, {\it Some Applications of Differential
Topology in General Relativity}, J. Geometry and Physics {\bf 13},
357-377 (1994).
\bibitem[Ch94b]{Ch94} A. Chamblin, {\it On the Obstructions to Non-Cliffordian
Pin Structures}, Commun. Math. Phys. {\bf 164}, 67-87, (1994).
\bibitem[Ch97]{Ch97} A. Chamblin, {\it On the Superselection Sectors of
Fermions}, preprint DAMTP R-97/4; hep-th/9704099, (1997).
\bibitem[Che54]{Che1} C. Chevalley, {\bf The Algebraic Theory of Spinors}
(Columbia University Press, New York, 1954).
\bibitem[Che55]{Che2} C. Chevalley, {\it The construction and study of certain
important algebras}, Publications of Mathematical Society of Japan No 1
(Herald Printing, Tokyo, 1955).
\bibitem[CF96]{CF97} J.S.R. Chisholm, R.S. Farwell, {\it Properties of
Clifford Algebras for Fundamental Particles}, in {\bf Clifford (Geometric)
Algebras}, ed. W. Baylis (Birkhuser, 1996), pp. 365--388.
\bibitem[CDD82]{CDD82} Y. Choquet-Bruhat, C. De Witt-Morette and
M. Dillard-Bleick, {\bf Analysis, Manifolds and Physics} (North-Holland
Publ. Co., Amsterdam, 1982).
\bibitem[Cliff]{3} W. K. Clifford, {\it Applications of Grassmann's
extensive algebra}, Amer. J. Math. {\bf 1}, 350, (1878).
\bibitem[Cor84]{Cor84} J.F. Cornwell, {\bf Group Theory in Physics}
(Academic Press, 1984).
\bibitem[Cru90]{Cru91} A. Crumeyrolle, {\bf Orthogonal and Symplectic
Clifford Algebras. Spinor Structures} (Kluwer Acad. Publ., Dordrecht, 1990).
\bibitem[D\c{a}b88]{Dab88} L. D\c{a}browski, {\bf Group Actions on Spinors}
(Bibliopolis, Naples, 1988).
\bibitem[DP86]{DP86} L. D\c{a}browski, R. Percacci, {\it Spinors and
Diffeomorphisms}, Commun. Math. Phys. {\bf 106}, 691-704, (1986).
\bibitem[DP87]{DP87} L. D\c{a}browski, R. Percacci, {\it Diffeomorphisms,
orientations, and pin structures in two dimensions}, J. Math. Phys. {\bf 29},
580-593, (1987).
\bibitem[DR89]{DR89} L. D\c{a}browski, M. Rinaldi, {\it Spin structures, spinors
and the Dirac operator: real versus complex manifolds}, J. Geometry and
Physics {\bf 6}, 651-656 (1989).
\bibitem[DT86]{DT86} L. D\c{a}browski, A. Trautman, {\it Spinor structures on
spheres and projective spaces}, J. Math. Phys. {\bf 27}, 2022-2088, (1986).
\bibitem[DW90]{DW90} B. De Witt, {\it The Pin Groups in Physics}, Phys.
Rev. {\bf D41}, p. 1901 (1990).
\bibitem[DWGK]{DWGK} C. De Witt-Morette, Shang-Jr Gwo, E. Kramer,
{\it Spin or Pin?}, preprint (1997); 
http://www.rel.ph.utexas.edu/Members/cdewitt/SpinOrPin1.ps.
\bibitem[FRO90a]{FRO90a} V.L. Figueiredo, W.A. Rodrigues, Jr., E.C. Oliveira,
{\it Covariant, algebraic, and operator spinors}, Int. J. Theor. Phys. {\bf 29},
371-395, (1990).
\bibitem[FRO90b]{FRO90b} V.L. Figueiredo, W.A. Rodrigues, Jr., E.C. Oliveira,
{\it Clifford algebras and the hidden geometrical nature of spinors},
Algebras, Groups and Geometries {\bf 7}, 153-198, (1990).
\bibitem[Fr97]{Fr97} Th. Friedrich, {\bf Dirac-Operatoren in der
Riemannschen Geometrie} (Vieweg-Verlag Braunschweig/Wiesbaden, 1997). 
\bibitem[Fr98]{Fr98} Th. Friedrich, {\it Cartan Spinor Bundles on Manifolds},
preprint SFB 288, N 303, TU-Berlin (1998).
\bibitem[FT96]{FT96} Th. Friedrich, A. Trautman, {\it Clifford structures
and spinor bundles}, preprint SFB 288, N 251, TU-Berlin, (1996).
\bibitem[FT99]{FT99} Th. Friedrich, A. Trautman, {\it Spin spaces,
Lipschitz groups, and spinor bundles}, preprint SFB 288, N 362, 
TU-Berlin (1999).
\bibitem[Ger68]{Ger68} R.P. Geroch, {\it Spinor Structure of Spacetimes in
General Relativity: I}, J. Math. Phys. {\bf 9}, 1739-1744 (1968).
\bibitem[GP82]{GP82} W. Greub, H.R. Petry, {\it Spinor structures},
Lect. Notes Math. {\bf 905}, 170-185 (1982).
\bibitem[Hae56]{Hae56} A. Haefliger, {\it Sur l'extension du groupe
structural d'un espace fibre}, C.R. Acad. Sci. Paris. {\bf 243}, 558-560
(1956). 
\bibitem[Hest66]{Hest66} D. Hestenes, {\bf Space-Time Algebra} (Gordon \&
Breach, New York, 1966).
\bibitem[Hest67]{Hest67} D. Hestenes, {\it Real spinor fields}, J. Math.
Phys. {\bf 8}, 798-808, (1967).
\bibitem[Hest76]{Hest76} D. Hestenes, {\it Observables, operators, and
complex numbers in the Dirac theory}, J. Math. Phys. {\bf 16}, 556-571,
(1976). 
\bibitem[Hest90]{Hest90} D. Hestenes, {\it The Zitterbewegung Interpretation
of Quantum Mechanics}, Found. Phys. {\bf 20}, 1213-1232 (1990).
\bibitem[Ish78]{Ish78} C.J. Isham, {\it Spinor fields in four dimensional
space-time}, Proc. R. Soc. Lond. A. {\bf 364}, 591-599 (1978).
\bibitem[Kar68]{Kar68} M. Karoubi, {\it Algebres de Clifford et K-theorie},
Ann. scient. Ec. Norm. Sup. $4^e$ ser. t.1, 161-270 (1968).
\bibitem[Kar79]{Kar} M. Karoubi, {\bf K-Theory. An Introduction} 
(Springer-Verlag, Berlin, 1979).
\bibitem[KT89]{KT89} R.C. Kirby, L.R. Taylor, {\it Pin Structures on
Low-Dimensional Manifolds}, London Math. Soc. Lecture Notes, no. 151, CUP,
(1989).
\bibitem[Lips]{Lips} R. Lipschitz, {\bf Untersuchungen \"{u}ber die Summen
von Quadraten} (Max Cohen und Sohn, Bonn, 1886).
\bibitem[Lou81]{Lou81} P. Lounesto, {\it Scalar Products of Spinors and an
Extension of Brauer-Wall Groups}, Found. Phys. {\bf 11}, 721-740, (1981).
\bibitem[Lou93]{Lou93} P. Lounesto, {\it Clifford algebras and Hestenes
spinors}, Found. Phys. {\bf 23}, 1203-1237 (1993).
\bibitem[Lou96]{Lou96} P. Lounesto, {\it Counter-Examples in Clifford
Algebras}, Advances in Applied Clifford Algebras {\bf 6}, 69-104 (1996).
\bibitem[LM89]{LM89} H.B. Lowson, M.-L. Michelsohn, {\bf Spin Geometry}
(Princeton University Press, Princeton 1989).
\bibitem[Mil63]{Mil63} J. Milnor, {\it Spin structures on manifolds},
Enseign. Math. {\bf 9}, 198-203 (1963).
\bibitem[Port]{Port} I. Porteous, {\bf Topological Geometry} 
(van Nostrand, London, 1969).
\bibitem[Ras55]{Rash} P.K. Rashevskii, {\it The Theory of Spinors},
(in Russian) Uspekhi Mat. Nauk {\bf 10}, 3-110 (1955); translation in
Amer. Math. Soc. Transl. (Ser.2) {\bf 6}, 1 (1957).
\bibitem[RRSV95]{RRSV95} D.L. Rapoport, W.A. Rodrigues, Jr., Q.A.G. de
Souza, and J. Vaz, Jr., {\it The Riemann-Cartan-Weyl geometry generated
by a Dirac-Hestenes spinor field}, Algebras, Groups and Geometries {\bf 11},
23-35 (1995).
\bibitem[RF90]{RF90} W.A. Rodrigues, Jr. and V.L. Figueiredo, {\it Real
spin-Clifford bundle and the spinor structure of the spacetime}, Int. J.
Theor. Phys. {\bf 29}, 413-424 (1990).
\bibitem[RO90]{RO90} W.A. Rodrigues, Jr. and E.C. Oliveira, {\it Dirac
and Maxwell equations in the Clifford and spin-Clifford bundles}, Int.
J. Theor. Phys. {\bf 29}, 397-412 (1990).
\bibitem[RS93]{RS93} W.A. Rodrigues, Jr. and Q.A.G. de Souza, {\it
The Clifford bundle and the nature of the gravitational field},
Found. Phys. {\bf 23}, 1465-1490 (1993).  
\bibitem[RVR93]{RVR93} W.A. Rodrigues, Jr., J. Vaz, Jr. and E. Recami,
{\it About Zitterbewegung and electron structure}, Phys. Lett. {\bf B 318},
623-628 (1993).
\bibitem[RSVL]{RSVL} W.A. Rodrigues, Jr., Q.A.G. de Souza, J. Vaz, Jr.,
P. Lounesto, {\it Dirac-Hestenes spinor fields in Riemann-Cartan spacetime},
Int. J. Theor. Phys., {\bf 35}, 1849-1900, (1996).
\bibitem[Sal81a]{Sal81a} N. Salingaros, {\it Realization, extension, and
classification of certain physically important groups and algebras},
J. Math. Phys. {\bf 22}, 226-232, (1981).
\bibitem[Sal81b]{6} N. Salingaros, {\it Algebras with three anticommuting
elements. II. Two algebras over singular field}, J. Math. Phys. {\bf 22}(10),
(1981).
\bibitem[Sal82]{7} N. Salingaros, {\it On the classification of Clifford
algebras and their relation to spinors in $n$ dimensions}, J. Math. Phys.
{\bf 23}(1), (1982).
\bibitem[Sal84]{Sal84} N. Salingaros, {\it The relationship between finite
groups and Clifford algebras}, J. Math. Phys. {\bf 25}, 738-742, (1984).
\bibitem[Sch49]{Sch49} J.A. Schouten, {\it On the geometry of spin spaces},
Indag. Math. {\bf XI}, 3,4,5, (1949).
\bibitem[Tr92]{Tr92} A. Trautman, {\it Spinors and the Dirac operator
on hypersurfaces. I. General Theory}, J. Math. Phys. {\bf 33}, 4011-4019,
(1992).
\bibitem[Var99a]{Var98a} V.V. Varlamov, {\it Generalized Weierstrass
representation for surfaces in terms of Dirac-Hestenes spinor field},
J. Geometry and Physics {\bf 32}(3), 241-251 (1999).
\bibitem[Var99b]{Var98b} V.V. Varlamov, {\it On spinor fields on the
surfaces of revolution}, Proc. Int. Conf. ''Geometrization of Physics IV'',
Kazan State University, Kazan, October 4--8, 1999, pp. 248--253.
\bibitem[VR93]{VR93} J. Vaz, Jr. and W.A. Rodrigues, Jr., {\it
Equivalence of the Dirac and Maxwell equations and quantum mechanics},
Int. J. Theor. Phys. {\bf 32}, 945-958 (1993).
\bibitem[Wh78]{Wh78} G.S. Whiston, {\it Compact Spinor Spacetimes},
J. Phys. A: Math. Gen. {\bf 11}, No. 7 (1978).
\end{thebibliography}
\end{document}